\documentclass{aa}

\usepackage{graphicx}
\usepackage{txfonts}
\usepackage{amsmath}    % Advanced maths commands
\usepackage{amssymb}    % Extra maths symbols
\usepackage{ulem}
\usepackage{hyperref}
\usepackage{xcolor}

\newcommand{\hi}{H{\sc{}i} }

\begin{document}

   \title{Toward Cosmicflows-4: The \hi data catalog}

   \author{A. Dupuy\inst{1}\thanks{dupuy@ipnl.in2p3.fr}
          \and
          H. M. Courtois\inst{1}
          \and 
          D. Guinet\inst{1}
          \and
          R. B. Tully\inst{2}
          \and
          E. Kourkchi\inst{2}
          }

   \institute{University of Lyon, UCB Lyon 1, CNRS/IN2P3, IUF, IP2I Lyon, UMR5822, 69622 Villeurbanne cedex, France\\
         \and
             Institute for Astronomy, University of Hawaii, 2680 Woodlawn Drive, Honolulu, HI 96822, USA\\
             }

   \date{Received XXX; accepted YYY}

\abstract
   {}
   {In this study, we present an update of a compilation of line width measurements of neutral atomic hydrogen (\hi) galaxy spectra at 21 cm wavelength. Our All Digital HI (ADHI) catalog consists of the previous release augmented with our new \hi observations and an analysis of archival data. This study provides the required \hi information to measure the distances of spiral galaxies through the application of the Tully-Fisher (TF) relation.}
   {We conducted observations at the Green Bank telescope (GBT) and reprocessed spectra obtained at the Nan\c{c}ay radiotelescope by the Nan\c{c}ay Interstellar Baryons Legacy Extragalactic Survey (NIBLES) and Kinematics of the Local Universe (KLUN) collaborations and we analyzed the recently published full completion Arecibo Legacy Fast ALFA (ALFALFA) 100\% survey in order to identify galaxies with good quality HI line width measurements.}
   {This paper adds new HI data adequate for TF use for 385 galaxies observed at GBT, 889 galaxies from archival Nan\c{c}ay spectra, and 1,515 rescaled Arecibo ALFALFA spectra. In total, this release adds 1,274 new good quality measurements to the ADHI catalog. Today, the ADHI database contains 18,874 galaxies, for which 15,433 have good quality data for TF use. The final goal is to compute accurate distances to spiral galaxies, which will be included in the next generation of peculiar velocities catalog: Cosmicflows-4.}
   {}


   \maketitle
%
%-------------------------------------------------------------------

\section{Introduction}

In 2006, we launched a large observational and theoretical program to study the large-scale structures and dynamics of the local Universe up to $z \sim 0.1$. Some remarkable outcomes of this project include the discovery of our home supercluster Laniakea \citep{2014Natur.513...71T}, the Dipole Repeller \citep[][DR]{Hoffman:2017aa}, the Cold Spot Repeller \citep{2017ApJ...847L...6C}, and the cartography of the Local Void \citep{2019ApJ...880...24T}.

Recently, we set up a distance-velocity calculator that is publicly accessible through the Extragalactic Distance Database\footnote{\url{edd.ifa.hawaii.edu}} (EDD).\ It is for any astronomer needing information on the local gravitational velocity field at a specified location in the local Universe \citep{2020AJ....159...67K}.

The \hi line widths data in our Cosmicflows catalogs come from our observations at Green Bank telescope (GBT) and our reprocessing of spectra obtained at other giant radio telescope archives such as Parkes, Nan\c{c}ay, Effelsberg, and Arecibo,  which were all gathered in the All Digital HI catalog \citep[ADHI, ][]{2009AJ....138.1938C, Courtois:2011ab, 2015MNRAS.447.1531C}. Table \ref{tab:adhi} shows a list of the various data and their respective publications in Column 2, which was used to build the ADHI catalog. Column 1 corresponds to the code used to refer to the source in the catalog.

The goal of this paper is to update the ADHI catalog, with 500 hours of observations at GBT (program GBT18A021) conducted in 2018 and 2019, with remeasured Nan\c{c}ay data from the Kinematics of the Local Universe (KLUN) and Nan\c{c}ay Interstellar Baryons Legacy Extragalactic Survey (NIBLES) projects \citep{Theureau:1998a,Theureau:2005aa,2007A&A...465...71T,Theureau:2017aa,van-Driel:2008aa,2016A&A...595A.118V} retrieved from public repositories. This paper also presents rescaled newly available xArecibo Legacy Fast ALFA (ALFALFA) Arecibo data. \citep{2018ApJ...861...49H}.

\begin{table}
\centering
\caption{All Digital HI catalog sources.}
\label{tab:adhi}
\begin{tabular}{p{0.4\columnwidth}p{0.5\columnwidth}}
\hline
Code & Literature source \\
\hline
ksk2004 & {\cite{2004AJ....128...16K}} \\
shg2005 & {\cite{2005ApJS..160..149S}} \\
hkk2005 & {\cite{2005A&A...435..459H}} \\
tmc2006 & {\cite{2006ASPC..351..429T}} \\
ghk2007 & {\cite{2007AJ....133.2569G}} \\
sgh2008 & {\cite{2008AJ....135..588S}} \\
kgh2008 & {\cite{2008AJ....136..713K}} \\
ctf2009 & {\cite{2009AJ....138.1938C}} \\
ctm2010-ctk2010 & {\cite{Courtois:2011ab}} \\
hgm2011 & {\cite{2011AJ....142..170H}} \\
archi12 & {\cite{Courtois:2015aa}} \\
hsm2013 & {\cite{2013MNRAS.432.1178H}} \\
\hline
tch2016 & {\cite{Theureau:2017aa}} \\
vbs2016 & {\cite{2016A&A...595A.118V}} \\
cdg2018 & This paper \\
\hline
\end{tabular}
\end{table}

\section{Measurement of the line width parameter}
\label{sec:measurew50}

In the Cosmicflows program, we measured the neutral hydrogen line width at 21cm wavelength enclosing 50\% of the cumulative HI line flux, $W_\mathrm{m50}$ \citep{Courtois:2011ab}. This parameter, $W_\mathrm{m50}$, is the line width measured at the flux level that is 50\% of the mean flux, averaged in channels within the wavelength range enclosing 90\% of the total integrated flux. However, the parameter $W_\mathrm{m50}$ is only an empirical measure of the true width of an HI galaxy velocity profile. A correction for redshift and instrumental broadening should be applied: $W^{c}_\mathrm{m50} = \frac{W_\mathrm{m50}}{1 +z} - 2\Delta v \lambda$, where $z$ is the redshift, $\Delta v$ is the smoothed spectral resolution, and $\lambda = 0.25$ is an empirically determined constant. The observed line width can also be adjusted by separating out the broadening from turbulent motions and offsets to produce an approximation to $2 V_\mathrm{max}$, where $V_\mathrm{max}$ characterizes the rotation rate over the main body of a galaxy. \cite{1985ApJS...58...67T} defined the parameter $W_\mathrm{mx}$ as:

\begin{equation}
\begin{aligned}
    W^2_\mathrm{mx}= & W^2_\mathrm{m50} + W^2_\mathrm{t,m50} \left[ 1 - 2 e^{-(W_\mathrm{m50}/W_\mathrm{c,m50})^2} \right] \\
    & - 2 W_\mathrm{m50} W_\mathrm{t,m50} \left[ 1 - e^{-(W\mathrm{m50}/W\mathrm{c,m50})^2} \right].
\end{aligned}
\label{eq:wmx}
\end{equation}

The parameters $W_\mathrm{c,m50}= 100$ km/s and $W_\mathrm{t,m50}= 9$ km/s were set after tests conducted in \citet{2009AJ....138.1938C}, and they characterized the transition from boxcar to Gaussian intrinsic profiles and the turbulent broadening for the observed line width considered, respectively. It was then related to the rotation rate $V_\mathrm{max}$ by:
\begin{equation}
    2 V_\mathrm{max} = \frac{W_\mathrm{mx}}{\sin{(i)}},
\end{equation}
where $i$ is the inclination of the galaxy from face-on relative to the observer. Inclinations have been evaluated using an online graphical tool, Galaxy Inclination Zoo (GIZ)\footnote{\url{http://edd.ifa.hawaii.edu/inclination/index.php}}, in a collaborative science project with citizens. Please refer to \S 2.3 of \citet{2020ApJ...902..145K} for further details. Details regarding the $W_\mathrm{m50}$ and $W_\mathrm{mx}$ line width parameters and comparisons with alternatives are discussed in \citet{2009AJ....138.1938C}. 

Finally, the error of the line width, $e_W$, is given by \citep{2009AJ....138.1938C}:
\begin{equation}
\begin{aligned}
    \mathrm{S/N} \geq 17 \;\;\;\; & e_W = 8; \\
    2 < \mathrm{S/N} < 17 \;\;\;\; & e_W = 21.6 - 0.8 \; \mathrm{S/N}; \\
    \mathrm{S/N} \leq 2 \;\;\;\; & e_W = 70 - 25 \; \mathrm{S/N}, \\
\end{aligned}
\end{equation}
where S/N is the signal-to-noise ratio. An HI target is considered adequate for estimating its distance with the Tully-Fisher (TF). 

Although the KLUN, NIBLES, and ALFALFA data have already been processed by their own collaborations, with methodologies described in their respective literature, these data samples have been reprocessed with the analysis described in this section in order to include them in the ADHI catalog, which gathers various HI data treated with that same procedure.

\section{Data}

\begin{figure*}
\centering
\includegraphics[width=1\textwidth]{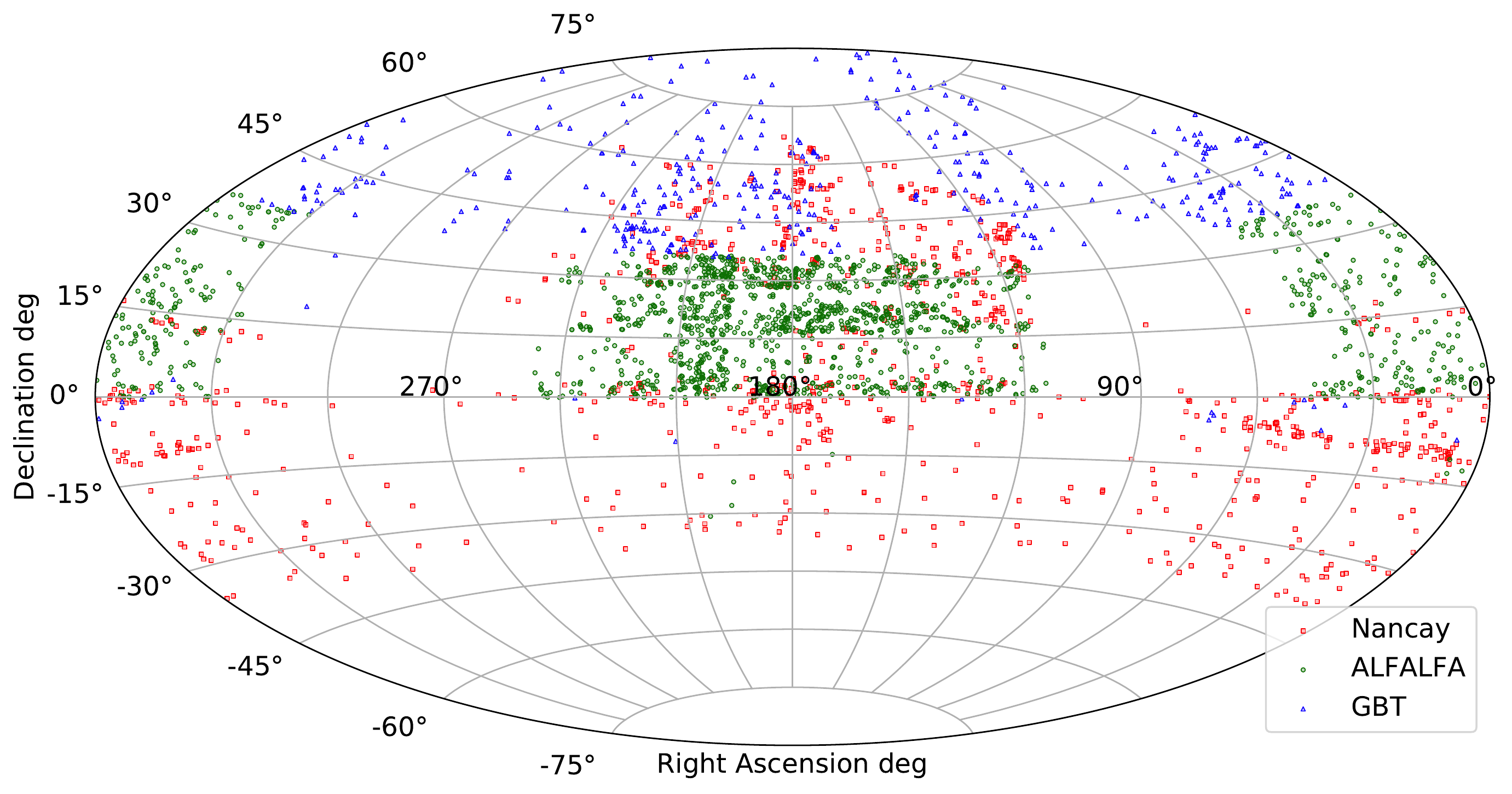}
\caption{Sky distribution in equatorial coordinates of all HI data considered in this paper. The red squares, green dots, and blue triangles correspond to the HI data from the Nan\c{c}ay (NIBLES$_I$ AND KLUN17 HI data releases), Arecibo (ALFALFA), and Greenbank telescopes, respectively. All of the galaxies shown are new additions to the ADHI catalog, and their line width measurements are adequate for the use of the TF relation.}
\label{fig:newHI_aitoff}
\end{figure*}

\begin{figure}
\includegraphics[width=0.5\textwidth]{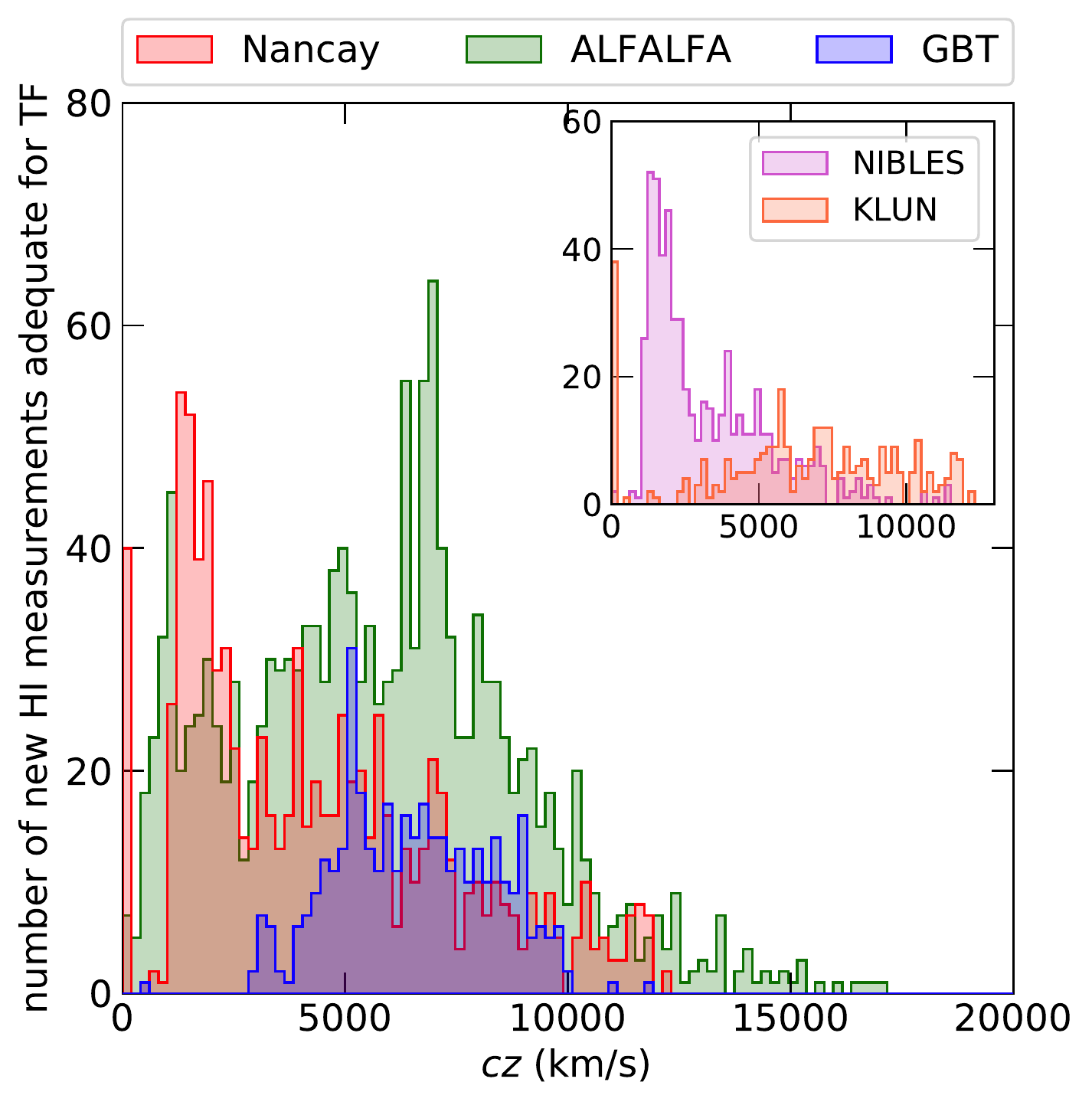}
\caption{Redshift distribution of new HI data described in this paper. The Nan\c{c}ay, Arecibo, and GBT data are shown in red, green, and blue, respectively. Only new additions to the ADHI catalog that are adequate for the use of the TF relation are considered. The subset plot in the top right corner shows the redshift distribution of Nan\c{c}ay data only. The NIBLES$_I$ and KLUN17 data are in purple and orange, respectively.}
\label{fig:histcz}
\end{figure}

\begin{figure*}
\centering
\includegraphics[width=0.9\textwidth]{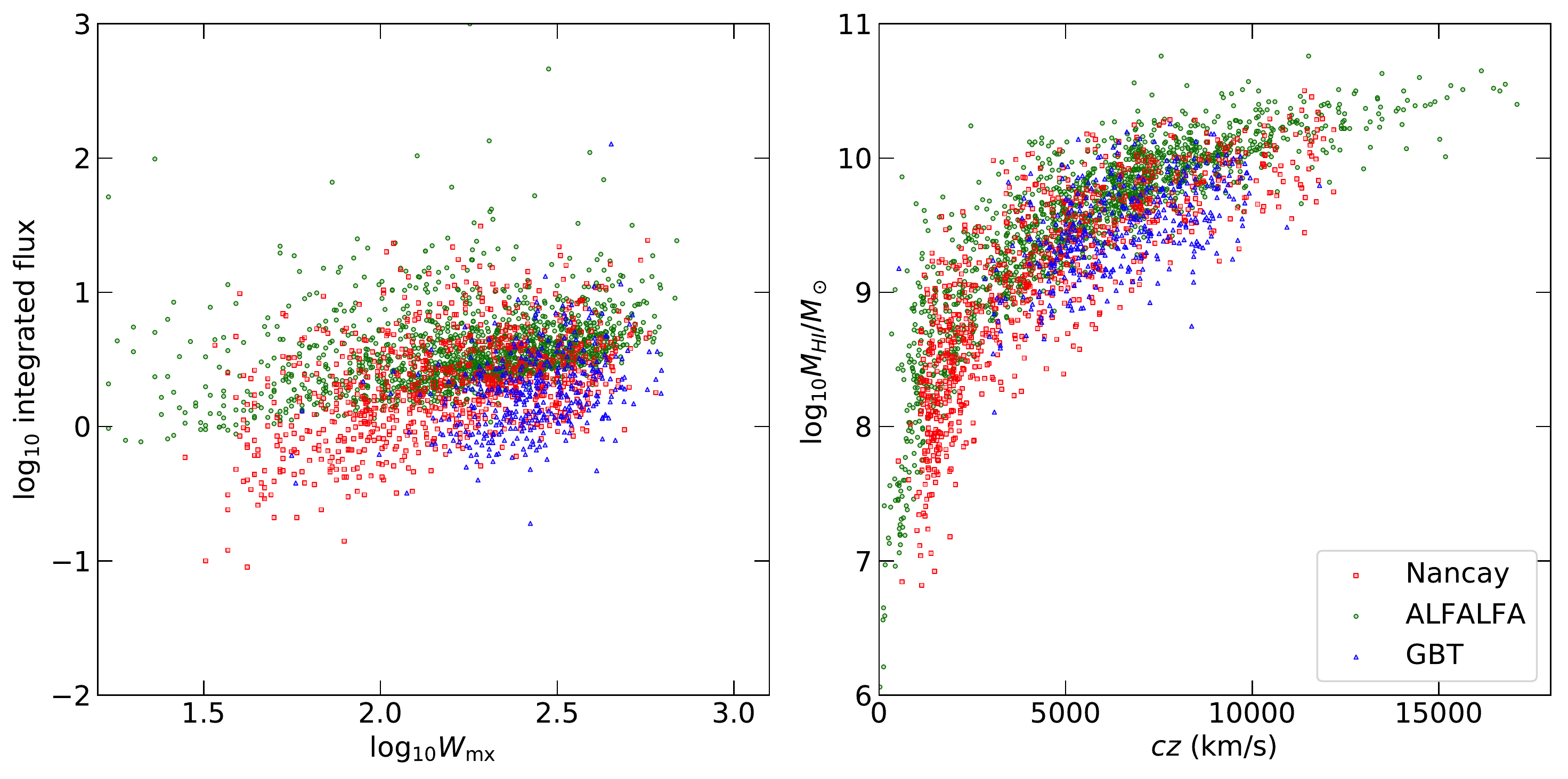}
\caption{{\bf Left:} Logarithm of the integrated flux as a function of the logarithm of the line width. {\bf Right:} Distribution of HI mass as a function of the redshift. Nan\c{c}ay data (NIBLES$_I$ and KLUN17) are represented by red squares, Arecibo data by green circles, and GBT data by blue triangles. Only new additions to the ADHI catalog that are adequate for the use of the TF relation are considered.}
\label{fig:completeness}
\end{figure*}

\subsection{Nan\c{c}ay: NIBLES$_I$ and KLUN17}

In this section, we consider two HI data releases that were obtained at the Nan\c{c}ay radio telescope by the KLUN and NIBLES collaborations: the KLUN17 release \citep{Theureau:2017aa} and the Nan\c{c}ay HI survey NIBLES$_I$ \citep{2016A&A...595A.118V}. The previous releases of KLUN data have been already included in the Cosmicflows catalogs, listed under the source code \texttt{tmc2006} in the ADHI catalog (see Table \ref{tab:adhi}).

Figure \ref{fig:newHI_aitoff} shows the sky distribution in equatorial coordinates of all HI data considered in this paper. The Nan\c{c}ay data considered in this section (both KLUN17 and NIBLES$_I$) are represented by red squares. The other samples displayed in this figure are described further below in the next sections. The data points in the Nan\c{c}ay samples are distributed homogeneously in the two regions.

The redshift distribution of the two Nan\c{c}ay samples is shown in red in Figure \ref{fig:histcz}. Similarly to Fig \ref{fig:newHI_aitoff}, the other HI samples are detailed later in the paper. We mainly find nearby galaxies, essentially below 6,000 km/s. The inset plot in the top right corner of Figure \ref{fig:histcz} shows the redshift distribution of each Nan\c{c}ay subset. The purple and orange histograms represent the NIBLES$_I$ and KLUN17 subsamples, respectively. The NIBLES$_I$ sample primarily contains nearby galaxies below 3,000 km/s, and very few distant galaxies up to 8,000 km/s. In contrast, the KLUN17 data slightly contribute to the addition of distant galaxies for the next Cosmicflows catalog, as it mostly contains galaxies above 5,000 km/s, and up to 12,000 km/s.

All data have been duly reduced and analyzed by their respective collaborations. However, for consistency, in order to include the line width measurements in the ADHI database, the HI parameter needs to be remeasured with the method used in the Cosmicflows collaboration. We reprocessed 1,864 HI spectra from the NIBLES$_I$ collaboration. A total of 1,443 line widths were detected and measured, including 808 new additions to the ADHI database, of which 565 are of an adequate quality for TF measurement. Regarding the KLUN17 sample, we obtained 500 detections after remeasureming the line width on 828 HI spectra. This provides 393 new additions to ADHI, of which 324 are of a suffient high quality to be used for TF purposes. 

The left panel of Figure \ref{fig:completeness} shows the logarithm of the integrated flux as a function of the logarithm of the line width, and the right panel shows the distribution of HI mass as a function of the redshift. Identically to the previous Figures \ref{fig:newHI_aitoff} and \ref{fig:histcz}, the Nancay data are represented by red squares, and the rest of the figures are described later in the paper.

\begin{figure*}
\centering
\includegraphics[width=0.47\textwidth]{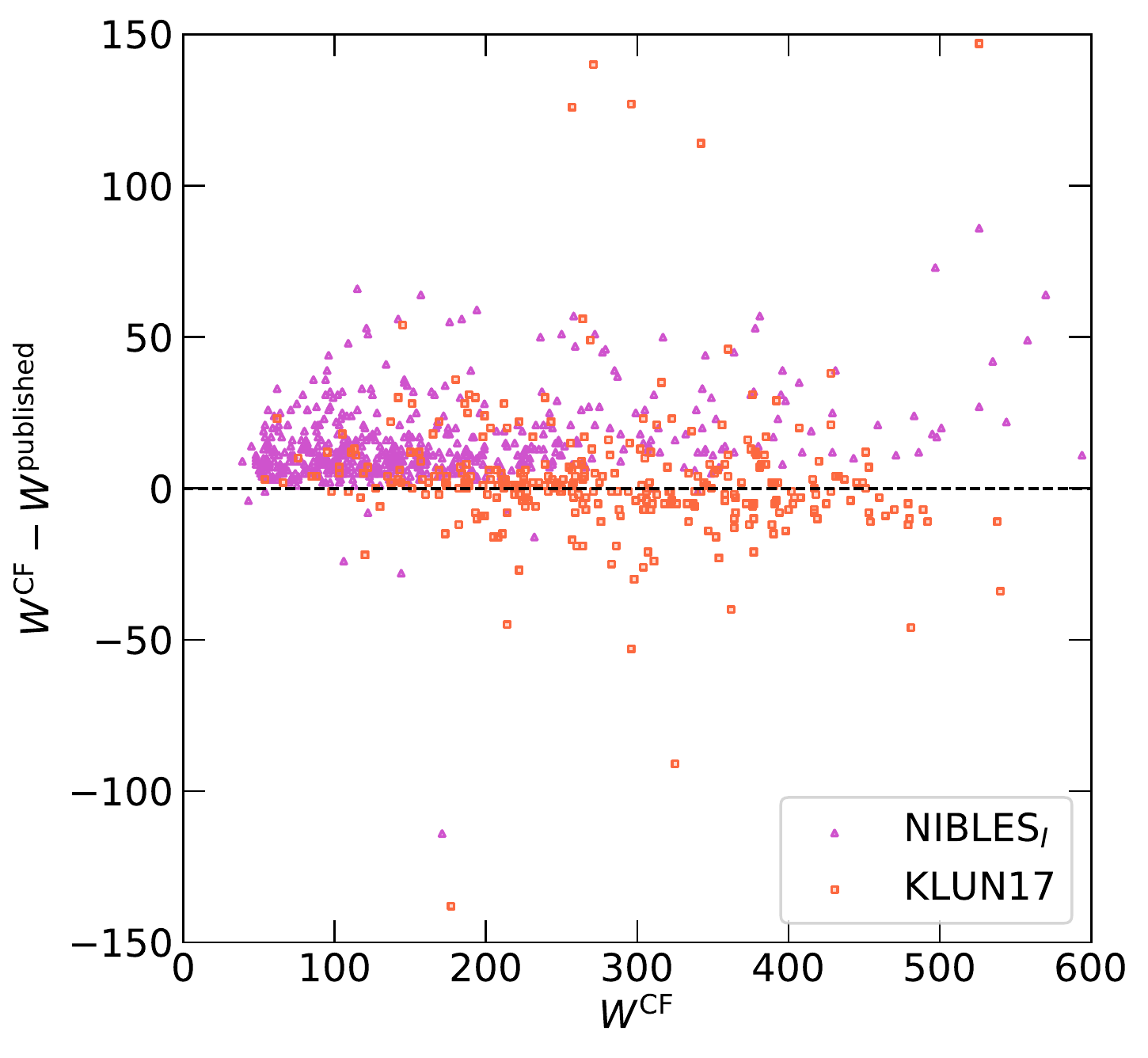}\includegraphics[width=0.47\textwidth]{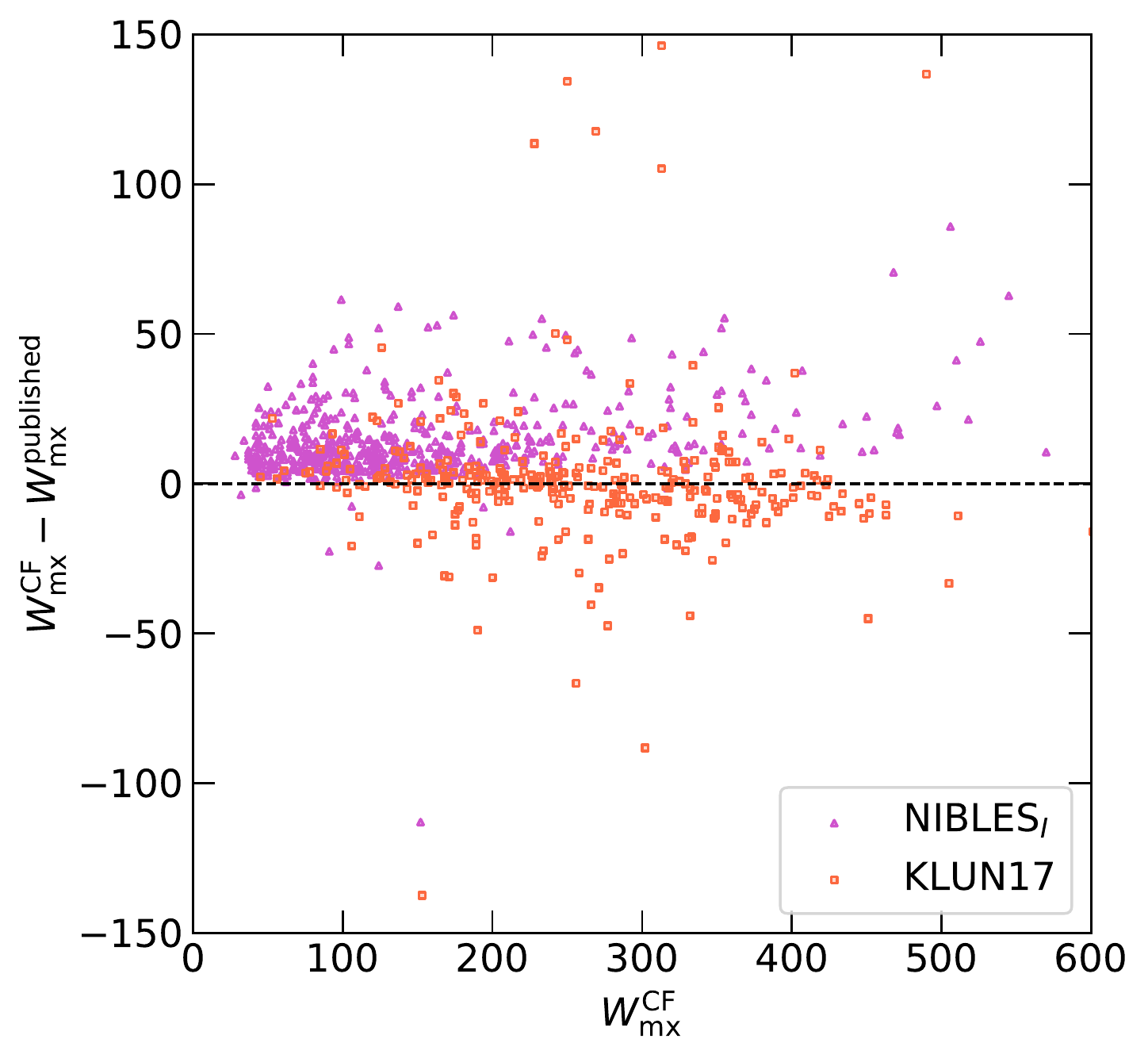}
\caption{Comparison of the measurements published in this paper and added to ADHI as $W^\mathrm{CF}$ to the ones published in the NIBLES$_I$ and KLUN17 data releases, represented by $W^\mathrm{published}$. The data obtained by the NIBLES$_I$ and KLUN17 collaborations are represented by purple triangles and orange squares, respectively. The black dotted line represents the case when $W^\mathrm{cf} = W^\mathrm{published}$. {\bf Left:} Comparison of the raw line widths $W^\mathrm{CF}_{50}$ and $W^\mathrm{published}_{50}$. {\bf Right:} Comparison of the corrected line widths $W^\mathrm{CF}_\mathrm{mx}$ and $W^\mathrm{published}_\mathrm{mx}$. In both panels, we observe a significant offset of 11 km/s for the NIBLES$_I$ sample.}
\label{fig:w50Nancay}
\end{figure*}

As discussed above, line widths were remeasured with the original Nan\c{c}ay spectra in order to be integrated into the ADHI catalog. In Figure \ref{fig:w50Nancay}, we compare the measurements published in this paper and added to ADHI as $W^\mathrm{CF}$ with the ones published by the NIBLES$_I$ and KLUN17 collaborations. These are indicated by $W^\mathrm{published}$ and represented as purple triangles and orange squares, respectively. The left panel shows the raw line widths $W_{50}$, while the right panel shows the line widths $W_\mathrm{mx}$ corrected with Eq. \ref{eq:wmx}. In both panels, the remeasured $W_{50}$ in the NIBLES$_I$ sample exhibits a significant positive offset of 11 km/s from the originally published $W_{50}$. This is explained by the difference in measuring the width at 50\% of the maximum flux and at 50\% of the mean integrated flux. The KLUN sample shows consistency with the Cosmicflows measurements for both raw and corrected line widths. 

\subsection{Arecibo: ALFALFA}

The Arecibo Legacy Fast ALFA (ALFALFA) survey is a blind extragalactic HI survey to conduct a census of the local HI universe over a cosmologically significant volume. 
The full 100\% ALFALFA extragalactic HI source catalog was published in \cite{2018ApJ...861...49H}. It contains more than 31,500 extragalactic  HI line sources detected out to z $<$ 0.06 in the Arecibo telescope's declination range of $0 < \delta < 38$ degrees.

\subsubsection{Identifying galaxies with adequate measurements}

The various HI parameters obtained by the ALFALFA collaboration have been published and are available online for 32,612 galaxies. However, to date, only 40\% of the actual HI spectra are available in a digital usable format (e.g., ascii, fits) \citep[$\alpha40$, 15,000 galaxies, published in ][]{2011AJ....142..170H}. These data were already remeasured by the Cosmicflows collaboration, bringing a total of 3,898 new good quality line width measurements to the ADHI catalog out of 15,000 reprocessed spectra \citep{2015MNRAS.447.1531C}. The source code \texttt{hgm2011} listed in Table \ref{tab:adhi} corresponds to the $\alpha40$ entry in the ADHI catalog. 
However, the remaining ALFALFA spectra of the full $\alpha100$ catalog were not available in the $\alpha40$ release, so the HI parameters cannot be directly re-estimated and included in ADHI.

In this section, we compare the parameters from the $\alpha40$ catalog obtained by ALFALFA and Cosmicflows. After establishing relations between these parameters, we then estimate the values of the HI parameters of the whole ALFALFA-100\% catalog, as if they were remeasured by the Cosmicflows collaborations. Throughout this section, parameters obtained by the ALFALFA (published in the $\alpha40$ catalog) and Cosmicflows collaborations are identified by the subscripts $\alpha$ and CF, respectively.

As our goal is to estimate which ALFALFA-100\% line width measurements may be adequate for deriving distances, we first consider the error, $e_W$, on the HI line width. The definition of this parameter differs between the two studies. When comparing the two parameters $e_W^\alpha$ and $e_W^\mathrm{CF}$, one can easily notice that $e_W^\alpha \neq e_W^\mathrm{CF}$.

The S/Ns, $\mathrm{S/N}^\mathrm{CF}$, and, $\mathrm{S/N}^\alpha$, are defined differently. In the Cosmicflows collaboration, the S/N is derived as the ratio of the signal at 50\% of the mean flux over the noise measured beyond the extremities of the signal \citep{2015MNRAS.447.1531C}. The S/N derived by the ALFALFA collaboration is defined in Eq. 2 of \cite{2011AJ....142..170H}. Both S/Ns, $\mathrm{S/N}^\mathrm{CF}$ and $\mathrm{S/N}^\alpha$, are compared directly in the left panel of Figure \ref{fig:alfalfa40}. Only galaxies adequate for TF, that is to say when $e_W^\mathrm{CF}\leq20$, are considered. The black dotted line represents the case when $\mathrm{S/N}^\mathrm{CF} = \mathrm{S/N}^\alpha$. One can relate the S/Ns from both collaborations by the following relation: 
\begin{equation}
\mathrm{S/N}^\alpha = 1.7 + 2.9 \; \mathrm{S/N}^\mathrm{CF},
\end{equation}
which is illustrated in the left panel of Figure \ref{fig:alfalfa40} by a solid red line. We can then estimate the value of $\mathrm{S/N}^\alpha$ in the Cosmicflows definition, which is denoted by $\mathrm{S/N}^{\alpha\rightarrow\mathrm{CF}}$ hereafter.

\subsubsection{Comparing the ALFALFA and Cosmicflows line widths}

The line widths obtained with the 70\% completion ALFALFA release ($\alpha70$) and Cosmicflows catalogs have been compared without reprocessing the spectra in \cite{2019ApJ...884...82K}, using a subsample of galaxies with adequate quality HI measurements.\ They were part of both catalogs at that time. The result was $W_\mathrm{mx}^\mathrm{CF} = W_{50}^\alpha - 6$, where $W_\mathrm{mx}^\mathrm{CF}$ is the line width in km/s taken from ADHI (including corrections, see equation \ref{eq:wmx}), and $W_{50}^\alpha$ is the line width provided by the ALFALFA $\alpha70$ catalog.

In this section, we follow the same methodology to compare line widths from both collaborations, considering now the newly published full 100\% ALFALFA catalog \cite{2018ApJ...861...49H} which contains 32,612 galaxies. Line widths extracted from this catalog are noted as $W_{50}^\alpha$. We are comparing them to a sample containing 15,433 good quality only line width measurements extracted from the ADHI dataset, corrected with equation \ref{eq:wmx} and represented by $W_{mx}^\mathrm{CF}$. The number of common galaxies in both datasets is 3,970.  

\begin{figure*}
\centering
\includegraphics[width=0.45\textwidth]{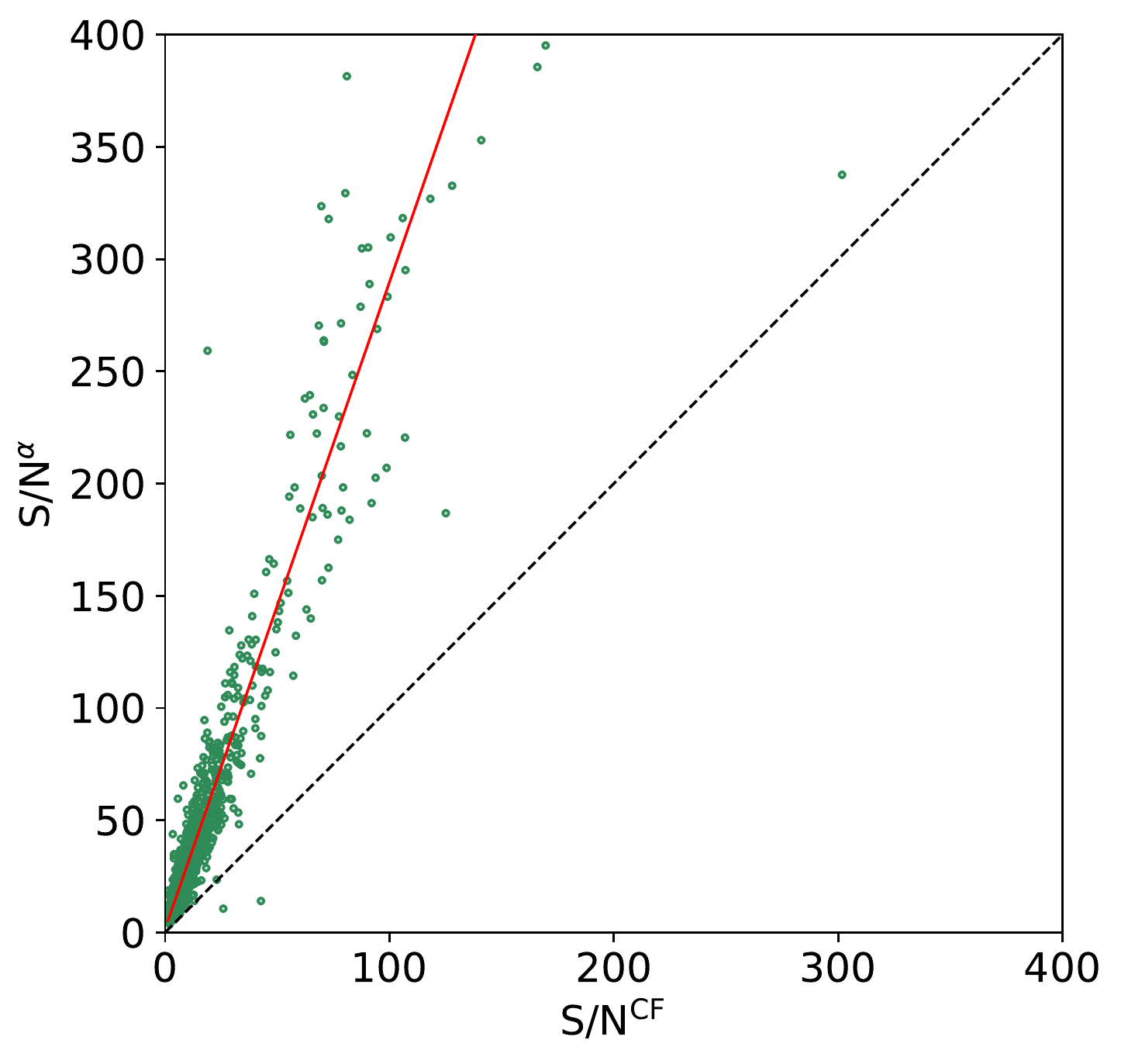}\includegraphics[width=0.45\textwidth]{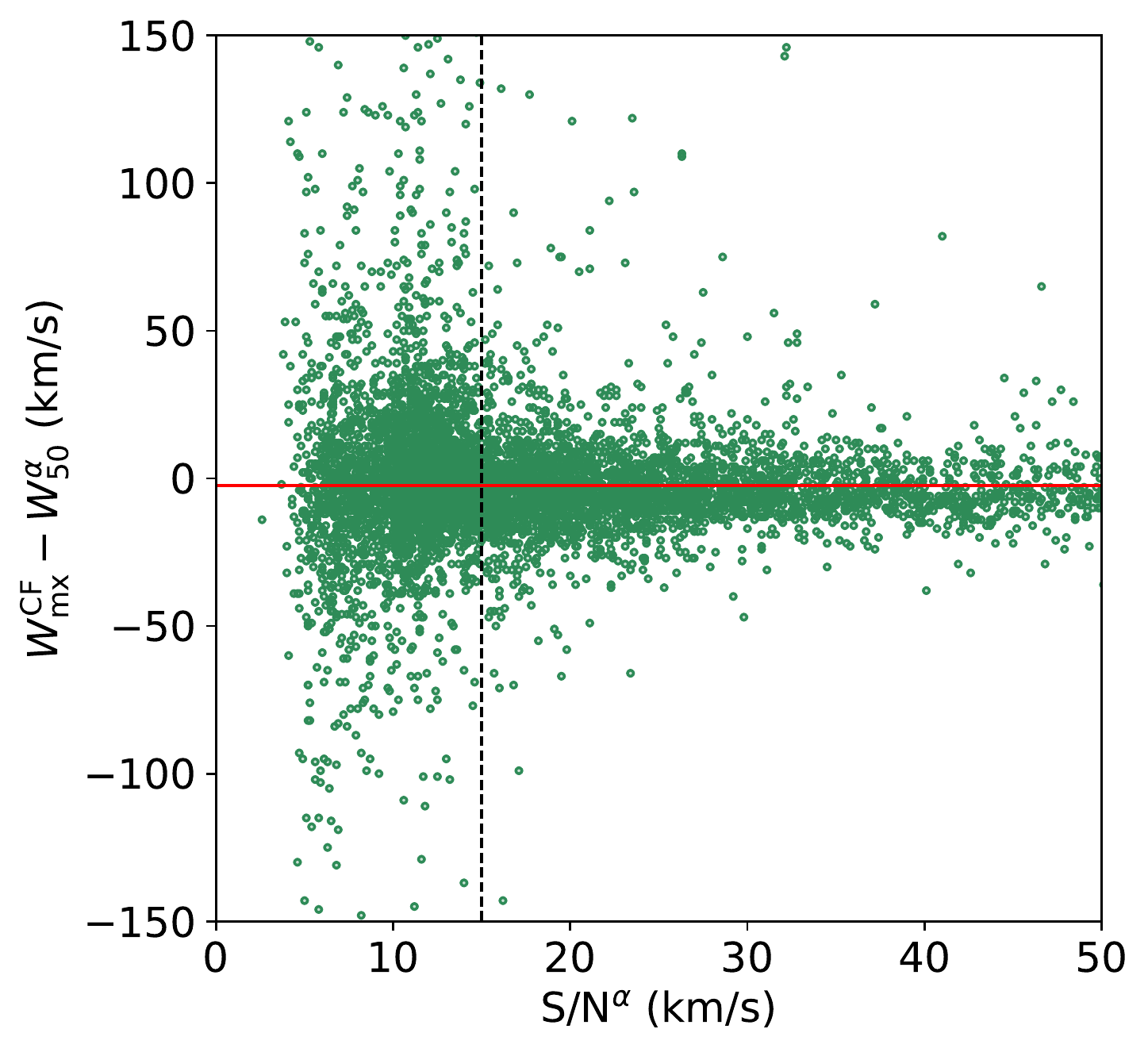}
\caption{Comparison of HI parameters derived by the Cosmicflows and ALFALFA collaborations identified by the subscripts CF and $\alpha$, respectively. Only galaxies available in the $\alpha$40 release are considered. {\bf Left:} Comparison of the S/Ns, $\mathrm{S/N}^\mathrm{CF}$ and $\mathrm{S/N}^\alpha$. {\bf Right:} Comparison of the line widths $W_\mathrm{mx}^\mathrm{CF}$ and $W_{50}^\alpha$. The difference $W_\mathrm{mx}^\mathrm{CF} - W_{50}^\alpha$ is represented as a function of the S/N, $\mathrm{S/N}^\alpha$. The black dotted line represents the threshold $\mathrm{S/N}^\alpha=15$ discussed in the text. Data above this threshold can be fitted to $W_\mathrm{mx}^\mathrm{CF} = W_{50}^\alpha - 2$ km/s, which is represented by a solid red line.}
\label{fig:alfalfa40}
\end{figure*}

In the right panel of Figure \ref{fig:alfalfa40}, we compare the line widths $W_{50}^\alpha$ and $W_{mx}^\mathrm{CF}$ by plotting the difference $W_{mx}^\mathrm{CF} - W_{50}^\alpha$ as a function of the S/N, $\mathrm{S/N}^\alpha$. We notice that there is significant dispersion at small S/Ns, where $\mathrm{S/N}^\alpha < 15$, which is represented by a black dotted line. Line width measurements above this threshold yield the following linear relationship between the two parameters:
\begin{equation}
W_\mathrm{mx}^\mathrm{CF} = W_{50}^\alpha - 2 ~(km/s)
\label{eq:alfa_conversion}
\end{equation}
represented in the right panel by a solid red line. This relation is slightly different from the one obtained previously in \cite{2019ApJ...884...82K}. We attribute this slight difference to the use of an earlier data release of the ALFALFA survey, $\alpha70$, by \cite{2019ApJ...884...82K}.

\subsubsection{Application to the full ALFALFA data}

In this section, equation \ref{eq:alfa_conversion} is applied to the full 100\% ALFALFA catalog (galaxies previously included in ADHI with high quality measurements are excluded) to compute how many and which galaxies could be added to the upcoming Cosmicflows-4 catalog. Galaxies with HI line width errors (in the Cosmicflows definition) of $e_W^{\alpha\rightarrow\mathrm{CF}} \leq 20$ km/s are suitable for the TF method and as such for inclusion in the new upcoming Cosmicflows-4 catalog. We identify 1,515 new galaxies which satisfy this criterion.

Figure \ref{fig:newHI_aitoff} displays the sky distribution of the additional 1,515 ALFALFA galaxies that are involved in the construction of the upcoming CF4 catalog.  ALFALFA galaxies are represented by green dots. A homogeneous coverage can be easily noticed in the region covered by Arecibo.

Figure \ref{fig:histcz} shows the redshift distribution of the new galaxies from ALFALFA in green. Three peaks are observed at 2,000 km/s, 5,000 km/s, and 8,000 km/s. This sample will thus be a significant contribution to CF4. It will include a large number of distant galaxies up to 17,000 km/s. However, it is important to note that only 338 galaxies ($\sim$22\% of the sample) have a redshift larger than 8,000 km/s.

The left panel of Figure \ref{fig:completeness} shows the integrated flux as a function of the line width. ALFALFA galaxies are represented by green dots. The blind ALFALFA survey mostly detects galaxies near the peak of the HI luminosity function where the largest volume is explored at a given flux.  Hence, detection is favored for giant galaxies with a large line width and intrinsic HI fluxes. Smaller galaxies at the same apparent flux levels are drawn from much smaller volumes. 

The distribution of the HI mass as a function of the redshift is shown in the right panel of Figure \ref{fig:completeness}. ALFALFA galaxies are once again displayed as green dots. The luminosity function (LF) of galaxies is defined as the number of galaxies per Mpc$^3$ in a luminosity interval $dM$ centered on magnitude $M$ \citep{2001MNRAS.327.1249Z}.
In  Figure \ref{fig:completeness}, we observe that the new galaxies observed satisfactorily with GBT are significantly less massive than the mean (or typical) ones from ALFALFA. The ALFALFA remeasurements are consistent in terms of the luminosity function with other surveys. The point concerning our GBT observations is that we do not aim to observe a representative survey of galaxies showing a good sample of a different luminosity or mass distribution galaxies. Our goal is to measure galaxies as far as possible, regarding  their velocity recession, in order to obtain an enlarged distribution of velocities with independent distances (using the TF extraction),  for our further studies of peculiar velocity. 
In other terms, we are not looking for a representative sample of galaxies following a luminosity function, rather we target galaxies with spatial, recession velocity and distances as various as possible.

\subsection{Green Bank telescope}

\subsubsection{Selection of targets}

Targets were selected from a compilation of two samples: a sample of flat galaxies from the Revised Flat Galaxy Catalog \citep[RFGC][]{Karachentsev:1999aa} to which a selection cut of $\delta>36\deg$ was applied to avoid overlap with the ALFALFA sample; and a sample of galaxies near to the Dipole Repeller \citep[][DR]{Hoffman:2017aa} -- galaxies were extracted from the LEDA database\footnote{\url{http://leda.univ-lyon1.fr/}} \citep{2014A&A...570A..13M} with the cuts $\alpha>20\deg$, $16<\delta<65\deg$, $3,000<cz<12,000$ km/s, and $i>50\deg$.

Several samples were added later to augment the number of potential targets. First, a sample of spiral galaxies extracted from LEDA with the following cuts applied: $\delta>36\deg$ (no overlap with ALFALFA), $3,000<cz<9,000$ km/s, and $i>55\deg$. The cuts were later modified in order to obtain more distant galaxies: $\delta>38\deg$, $3,000<cz<10,000$ km/s, and $40<i<85\deg$. Lastly, a few SNIa hosts from SNfactory \citep{2018arXiv180603849R,2002SPIE.4836...61A} with a redshift up to 12,000 km/s were added to the list of targets. 

All galaxies that are already part of the ADHI catalog with adequate HI profiles were removed from the samples described above. Several galaxies with almost adequate HI profiles from the GB300 and Nan\c{c}ay telescopes have also been added to the list of potential targets.

This extensive list now contains 5,460 galaxies, all of which have been inspected and are observed if they satisfy the following criteria. First, we needed to make sure that the galaxies were spiral galaxies with HI gas. One can look at the Pan-STARRS \citep{2016arXiv161205560C} optical images in the $g,r,i,z,$ and $y$ bands. A galaxy with HI, especially its disk, is hardly visible in the $y$ band, while it is very bright at the $g$ band. We also searched the optical spectra from SDSS DR12\citep{2015ApJS..219...12A} or LAMOST \citep{Luo_2015} for a significant H$\alpha$ emission line, hinting at the presence of young stars and therefore HI gas. Spectra containing Na, Mg, and Ca emission lines are indicative of old stellar populations and little HI. Available Pan-STARRS images and SDSS or LAMOST spectra were systematically inspected by eye for each target. A total of 54\% of the targets had an optical spectrum available. This ratio is not entirely satisfactory, but we did not find a recent, available optical spectrum for all our targets. Secondly, we did not observe collisions or interacting galaxies. If the distribution of the HI gas is disturbed, it does not allow us to measure the rotation speed of each galaxy.Moreover, the size of the radio lobe must be considered. Two galaxies within the radio lobe and located at the same redshift would lead to a confused spectrum because the galaxies spectra would overlap. Last but not least, we also checked the photometry quality and the inclination of the potential target to ensure the TF relation is applicable. Inclinations are indispensable to de-project the HI line widths. We require that inclinations be greater than 45 degrees from face-on. 

After inspection of 5,460 potential targets, 628 galaxies were selected and observed. The target selection procedure was not completely carried out in the sense that we observed well-defined targets with our first selection of criteria but which ultimately do not give a satisfactory result after observation.

\subsubsection{Observations: Strategy and planning}

A total of 610 hours of observations have been conducted from December 2017 to May 2019. During observations, the telescope was configured as follows. We used the L-band receiver \texttt{Rcvr1\_2}, detecting frequencies between 1.15 GHz and 1.73 GHz. For this receiver, the gain of the GBT is 2 K/Jy, and the size of the radiolobe is 9 arcmin. The spectral resolution considered during observations is 0.9 km/s.

The following ON-OFF-ON methodology was used throughout our observations. This observing strategy allows two ON-OFF scans to be conducted on the same source during a single session of observation. It consists of one 300 s scan on the source (ten integrations of 30 s), followed by a 300 s scan on the sky, then a second 300 s scan on the source. A total of 15 minutes is spent on a single source during a session. This strategy permits an additional 5 minutes of observation time per source and per session compared to the usual ON-OFF methodology.

During an observation run, targets are selected automatically from the list of inspected targets based on their proximity relative to the telescope orientation. Other target selection criteria are $0<\delta<85\deg$ and that the angular separation between the target and the Sun should be higher than 10 $deg$. After each run, we reduced the collected data as described in Section \ref{sec:datareduction} below and updated the target list, accordingly. Keywords are used to label and identify galaxies to be observed. The ON-OFF-ON cycles are repeated until the signal is sufficient to give an adequate line width measurement (S/N high enough to give an acceptable error on the line width, see Section \ref{sec:measurew50}). The total integration time per source goes from 15 min (one ON-OFF-ON cycle), to several hours for the most distant galaxies, or when the spectra were contaminated by radio frequency
interferences (RFIs) as detailed in Section \ref{sec:datareduction} below.

\subsubsection{Data reduction and RFIs}
\label{sec:datareduction}

Each integration of a single ON-OFF pair is calibrated with the \texttt{getsigref} function of \texttt{GBTIDL}\footnote{\url{http://gbtidl.nrao.edu}} in order to obtain the final spectra. After removing eventual RFIs as described in this section below, a Hanning smoothing (\texttt{hanning} function) is applied in order to obtain a resolution of 3.6 km/s.

In total, 628 HI spectra of galaxies have been obtained at GBT, of which 407 correspond to detections and 385 are acceptable for the use of the TF relation. The raw HI parameters measured on the 628 spectra obtained are available in Table \ref{tab:hiGBT}, as well as complementary data. Column 1 corresponds to the ID number of the target in the Principal Galaxy catalog (PGC). Column 2 gives an alternative name. Columns 3 and 4 correspond to the input (heliocentric) velocity $V_\mathrm{hel}$ and the velocity measured on the profile $V_\mathrm{50}$. The line width $W_\mathrm{50}$ and its uncertainty $e_{W}$, as well as the S/N and the measured HI line flux are provided in Columns 5, 6, 7, and 8, respectively. We note that we use the error codes $e_W = 100$ km/s and $e_w = 500$ km/s for confused spectra and non-detections, respectively. The inclination is listed in Column 9, and the total B-band magnitude is provided in Column 10. All non-HI parameters (input velocity, inclination, and magnitude) have been extracted from the LEDA database. All HI profiles and corresponding PanSTARRS optical images of the 628 galaxies observed can be found in Table \ref{tab:pngprofiles}, which is available electronically at the journal's homepage. 

\begin{table*}
\centering
\caption{Raw HI parameters measured on all 628 spectra obtained at GBT as well as complementary data. The full table is available electronically at the journal's homepage. The error $e_W = 100$ km/s corresponds to the confusion case where, for example, two galaxies are in the detection field, $e_W = 500$ km/s, which corresponds to the no detection case.}
\label{tab:hiGBT}
\begin{tabular}{cccccccccc}
\hline
PGC & Name & $V_\mathrm{hel}$  & $V_\mathrm{50}$ & $W_\mathrm{50}$ & $e_{W}$ & S/N & Flux & $i$ & $b_t$  \\
\hline
    676  &    UGC00085      &   5153    &   5157    &   277     &  16     &  6.9      &  3.7    &  48    &  14.46 \\
  22698 &     UGC04205    &     6850    &   6844    &   390   &    18   &    4.1  &      3.1   &   90   &   15.22 \\
  20121 &     UGC03655   &      6170    &   6056    &   201  &     49  &     0.8   &     1.5  &    42   &   14.49 \\
   9432 &     IC1799      &     5022    &   4926    &   422  &    100   &    6.9   &     6.6   &   64   &   14.89 \\
   8177 &     UGC01628    &     5556    &   5650    &    76   &   500   &    0.9   &     0.3   &   62   &   15.65 \\
\hline
\end{tabular}
\end{table*}

\begin{table*}
\centering
\caption{HI profiles and PanSTARRS multi-band images of the 628 galaxies observed at GBT. Columns 1 and 3 give the PGC number, the heliocentric velocity $V_\mathrm{hel}$ in km/s, and the HI profile after the measurement of the line width. In the case of non-detection, the raw spectra obtained from the GBT is shown in red. Columns 2 and 4 give the raw line width $W_{50}$ and its error $e_W$ in km/s, as well as the PanSTARRS image composite of the bands y/i/g. The full table is available electronically at the journal's homepage.}
\label{tab:pngprofiles}
\begin{tabular}{cccc}
\hline
PGC / $V_\mathrm{hel}$  & $W_\mathrm{50}$  / $e_W$  & PGC / $V_\mathrm{hel}$  & $W_\mathrm{50}$  / $e_W$  \\
Profile & PanSTARRS & Profile & PanSTARRS \\
\hline
PGC0000676  /  5157  &  277  /  16  &  PGC0022698  /  6844  &  390  /  18  \\
\includegraphics[width=4cm,trim={2cm 2cm 2cm 2cm},clip]{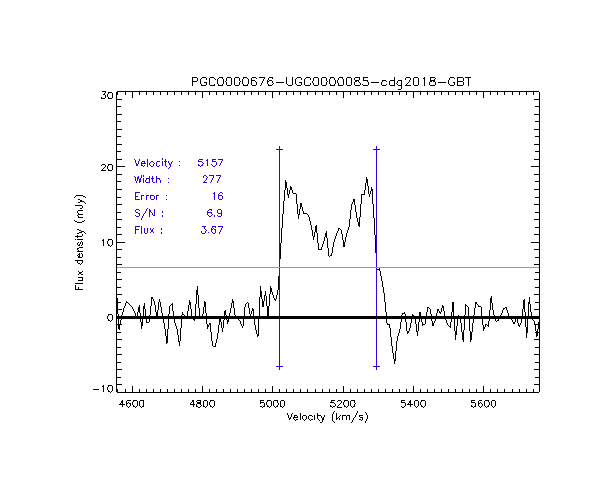} & \includegraphics[width=3cm]{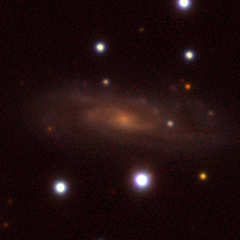} & \includegraphics[width=4cm,trim={2cm 2cm 2cm 2cm},clip]{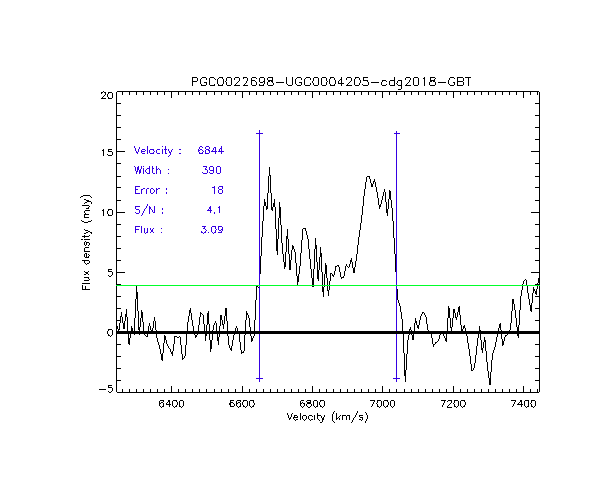} & \includegraphics[width=3cm]{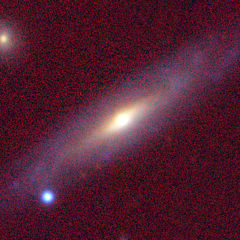} \\
\hline
PGC0020121  /  6056  &  201  /  49  &  PGC0009432  /  4926  &  422  /  100  \\
\includegraphics[width=4cm,trim={2cm 2cm 2cm 2cm},clip]{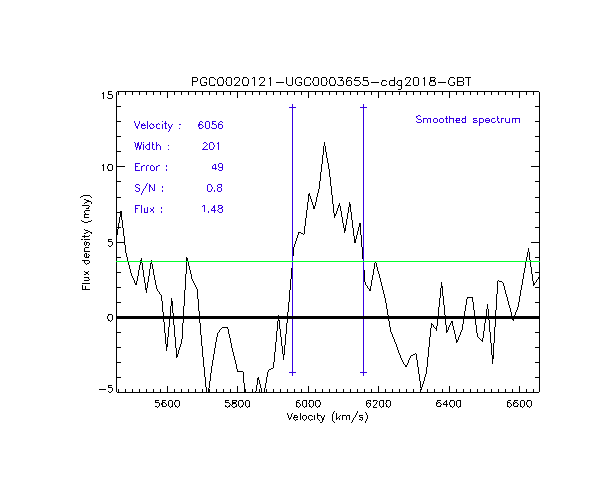} & \includegraphics[width=3cm]{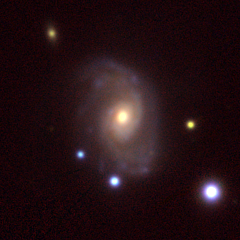} & \includegraphics[width=4cm,trim={2cm 2cm 2cm 2cm},clip]{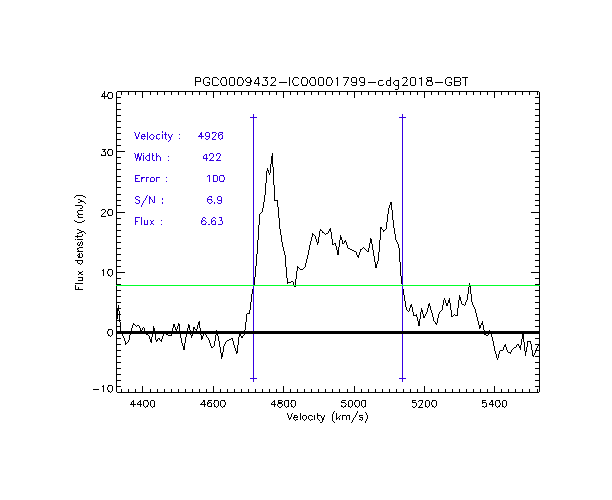} & \includegraphics[width=3cm]{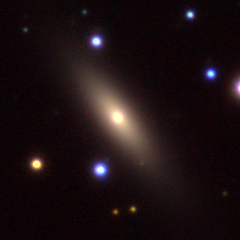} \\
\hline
PGC0008177  /  5650  &  76  /  500  &  &  \\
\includegraphics[width=4cm,trim={2cm 2cm 2cm 2cm},clip]{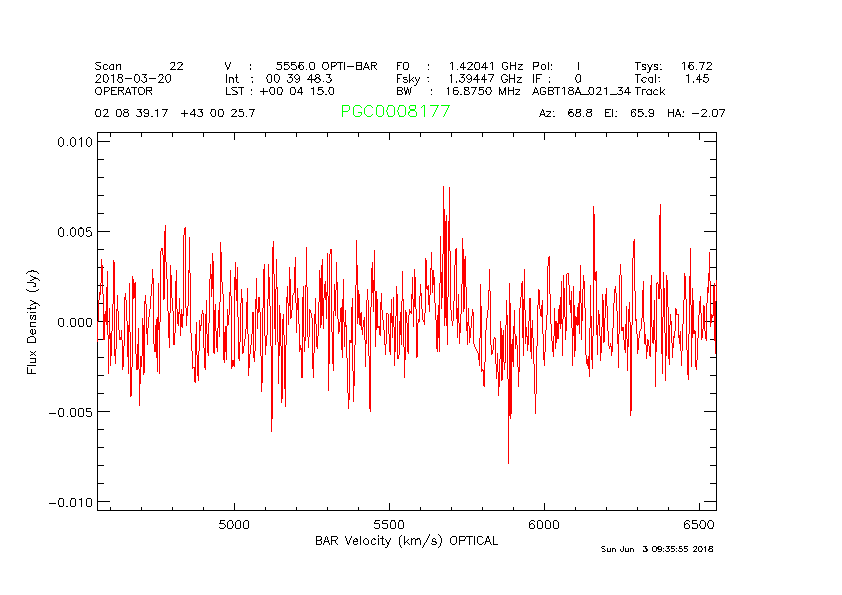} & \includegraphics[width=3cm]{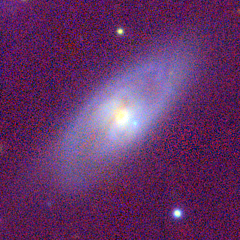} &  &  \\
\hline
\end{tabular}
\end{table*}

The main problems faced during data reduction arose from RFI. The single peak RFIs are removed easily by interpolation and do not affect the final spectra and measurements of HI parameters. 

\begin{figure}
\includegraphics[width=0.5\textwidth]{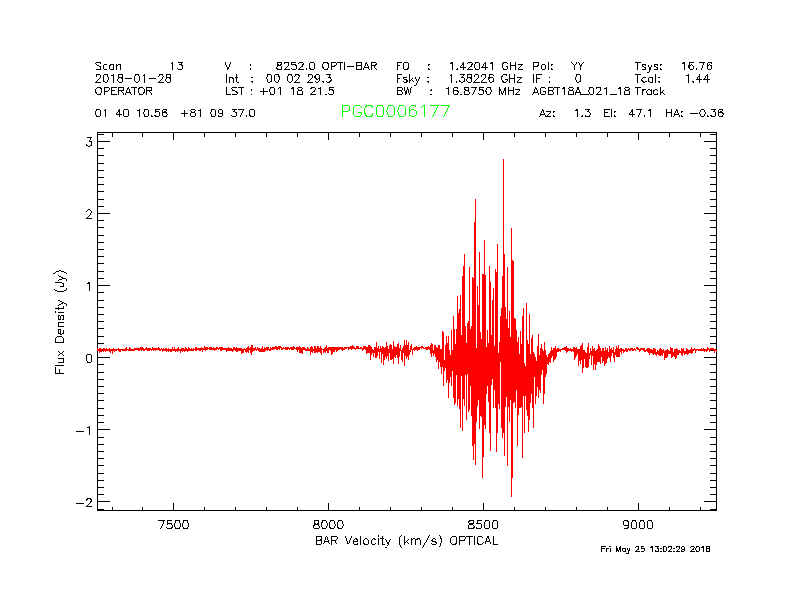}
\caption{HI spectra of PGC6177 located at $cz = 8,252$ km/s. A strong radio frequency interference (RFI) occurring frequently at 8,500 km/s can be noticed, while the HI line is not visible.}
\label{fig:gbt_rfitypes}
\end{figure}

However, a significant RFI located at 8,500 km/s (or 1.381 MHz) has been encountered frequently. It may be generated by a radar or a GPS system used for nuclear tests. This RFI destroyed a significant amount of HI spectra of distant galaxies located between 8,000 km/s and 9,000 km/s. The RFI is visible, for example, on the HI spectra of PGC6177 located at $cz = 8,252$ km/s, as shown in Figure \ref{fig:gbt_rfitypes}. The HI line on contaminated spectra is not visible at all as the amplitude of the RFI (up to $\approx10$ Jy) is much larger than the usual amplitude of an HI line ($\approx15$ mJy). We chose to ignore integrations of a single scan contaminated by this RFI when deriving the final HI spectra. The FFT methodology (sigma clipping in Fourier space) suggested by \cite{2016AJ....152...30H} has been tested on contaminated spectra with no success. Please refer to this publication for more details on the method. 

Finally, 192 targets have been observed for which no HI lines have been detected, and 29 spectra are confused. Sadly, a significant amount of observing time has been lost on these targets,  even though all targets have been deeply inspected prior to observing to check if an HI disk may be present. 

\subsection{New additions to the ADHI collection}

The sky distribution of the 385 new additions to the ADHI collection is shown in Figure \ref{fig:newHI_aitoff}. GBT galaxies are displayed as blue triangles. We focused on the northern celestial hemisphere to address the lack of the data coverage by Cosmicflows-3 in this region.   

The redshift distribution of these galaxies is represented in blue in Figure \ref{fig:histcz}. Mostly distant galaxies (up to 10,000 km/s) have been observed and added to ADHI.

The left panel of Figure \ref{fig:completeness} shows the log of the integrated flux as a function of the log of the HI line width. We mostly observed massive galaxies with low flux. The right panel shows the distribution of the HI mass as a function of the redshift. 

\begin{figure*}
\centering
\includegraphics[width=1\textwidth]{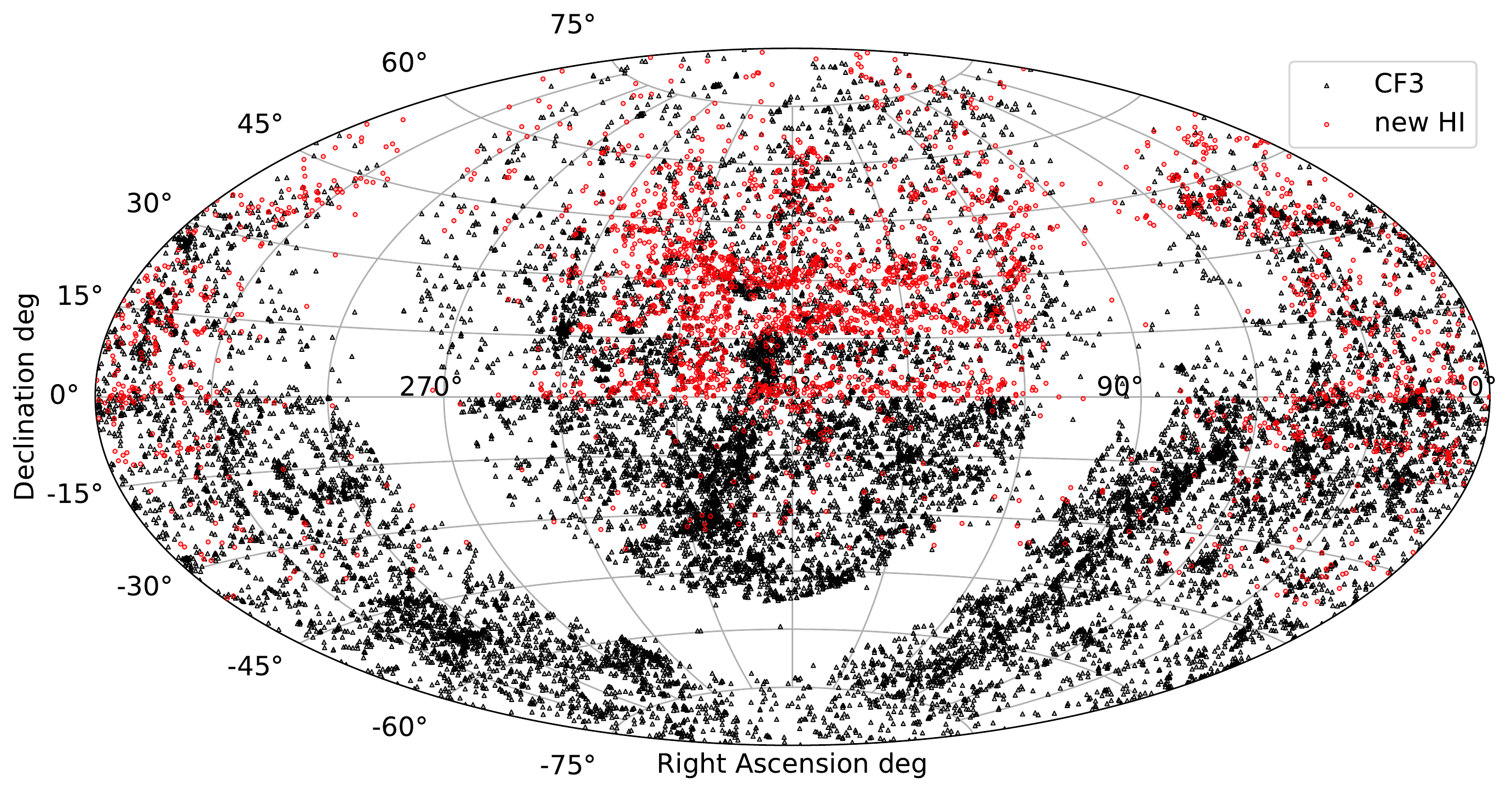}
\caption{Sky distribution in equatorial coordinates of all new line width measurements presented in this paper, which are adequate for TF distance measurement. The black triangles correspond to the Cosmicflows-3 catalog, while the red dots represent the new additional HI data.}
\label{fig:cf4_aitoff}
\end{figure*}

\section{From Cosmicflows-3 to Cosmicflows-4}

The sky distribution of all new line width measurements adequate for TF is shown in Figure \ref{fig:cf4_aitoff}. The black triangles correspond to the CF3 catalog, while the red dots represent the new HI data. Most of the new data are located in the north in order to correct the north-south asymmetry in Cosmicflows-3. The imbalance in the number of galaxies is indeed slightly corrected. In CF3, $\sim$70\% of the data were located in the south and only $\sim$30\% in the north. After adding the new HI data presented in this paper, $\sim$55\% of data should be in the south and $\sim$45\% in the north.

\begin{figure}
\includegraphics[width=0.47\textwidth]{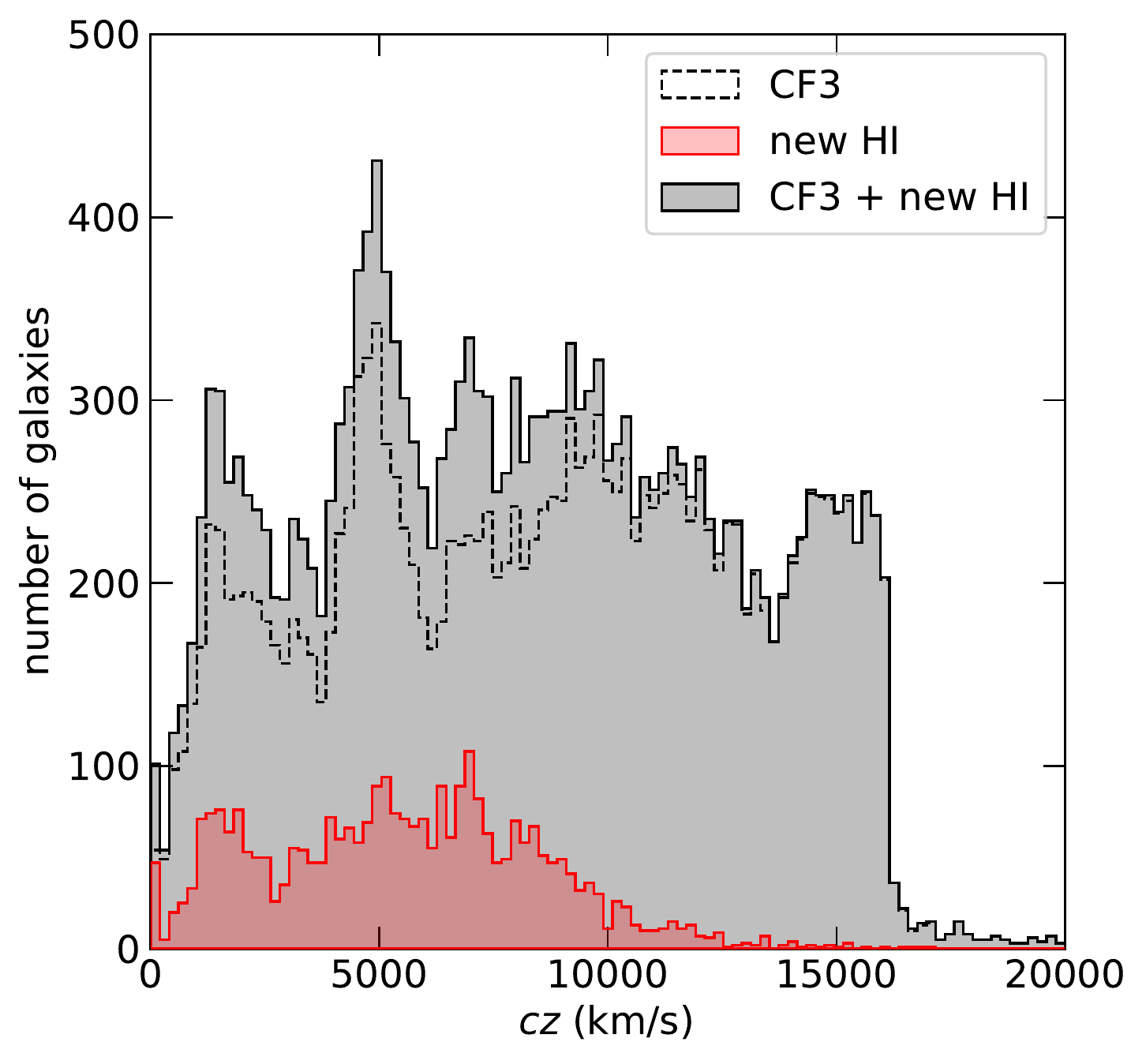}
\caption{Redshift distribution of CF3 data (black dotted line) and resulting addition into CF4 catalog (merging of CF3 and new HI data, gray with solid black line). The redshift distribution of the new HI data is shown in red.}
\label{fig:cf4_redshift}
\end{figure}

In Figure \ref{fig:cf4_redshift}, we compare the redshift distributions of the CF3 data and the new HI data to be added to CF4. The CF3 data are represented in gray with a black dotted line, while the new HI data are over-plotted and shown in red. The combination of CF3 with these new data to produce CF4 is shown in Figure \ref{fig:cf4_redshift} in gray with a solid black line. The contribution of the TF method to CF4 becomes less significant as $cz$ increases. The new HI data added in the north are not distant enough to fully compensate for the lack of data in this region, compared to the south which is mainly covered by the fundamental plane method with 6dF. 

We have updated the ADHI catalog by incorporating all galaxies with new HI observations at GBT and those with reprocessed Nan\c{c}ay HI measurements presented in this study, as listed in Table \ref{tab:newHI}. Column 1 corresponds to the ID number of the target in the Principal Galaxy catalog (PGC). An alternative name is given in Column 2. The source of the spectra and the telescope used to conduct the observation are given in Columns 3 and 4, respectively. The heliocentric velocity $V_\mathrm{hel}$ is provided in Column 5. Main HI parameters are listed in Columns 6, 7, and 8.

\begin{table*}
\centering
\caption{Newly observed and reprocessed HI measurements presented in this paper. The full table is available electronically at the journal's homepage.}
\label{tab:newHI}
\begin{tabular}{cccccccc}
\hline
PGC & Name & Source & Telescope & $V_\mathrm{hel}$ & $W_\mathrm{mx}$ & $e_{W}$ & S/N  \\
\hline
14330   & MCG-03-11-007 & {\cite{Theureau:2017aa}}     & Nan\c{c}ay  & 7536  & 376   & 18    & 3.3   \\ %klun
25180   & MCG+03-23-021 & {\cite{Theureau:2017aa}}     & Nan\c{c}ay  & 6150  & 194   & 17    & 5.7   \\ %klun
1862    & MCG-02-02-033 & {\cite{2016A&A...595A.118V}} & Nan\c{c}ay  & 5351  & 266   & 18    & 4.2   \\ %nibles
43634   & UGC08012      & this paper               & GBT         & 8160  & 290   & 18    & 3.5   \\ %gbt
47833   & MCG+06-30-042 & this paper               & GBT         & 7005  & 196   & 17    & 5.4   \\ %gbt
\hline
\end{tabular}
\end{table*}

\section{Conclusion}

We have acquired 385 high quality HI measurements from 628 targets using the GBT and 889 good measurements of the HI line width from 2,692 remeasured Nan\c{c}ay spectra. These 1,274 newly observed and reprocessed galaxies have been added to our ADHI catalog maintained at the Extragalactic Distance Database (EDD). Furthermore, we identify 1,515 new ALFALFA galaxies with spectra that may be sufficient for TF distance measurement. In total, this paper brings an additional 2,789 new galaxies that were not previously in our All Digital Catalog.

However, this study does not fully counterbalance the celestial north-south asymmetrical coverage of the HI observations in the Cosmicflows catalogs. In the north, future HI surveys, such as Apertif in the Netherlands and FAST in China, could potentially improve this imbalance.

\begin{acknowledgements}
HC is grateful to the Institut Universitaire de France and CNES for its support. AD acknowledges financial support from the Project IDEXLYON at the University of Lyon under the Investments for the Future Program (ANR-16-IDEX-0005). We acknowledge the use of the HyperLeda database. This project is partly financially supported by Region Rhone-Alpes-Auvergne.
\end{acknowledgements}


  \bibliographystyle{aa} % style aa.bst
  \bibliography{references} % your references Yourfile.ib

\begin{thebibliography}{37}
\expandafter\ifx\csname natexlab\endcsname\relax\def\natexlab#1{#1}\fi

\bibitem[{{Alam} {et~al.}(2015){Alam}, {Albareti}, {Allende Prieto}, {Anders},
  {Anderson}, {Anderton}, {Andrews}, {Armengaud}, {Aubourg}, {Bailey}, {Basu},
  {Bautista}, {Beaton}, {Beers}, {Bender}, {Berlind}, {Beutler}, {Bhardwaj},
  {Bird}, {Bizyaev}, {Blake}, {Blanton}, {Blomqvist}, {Bochanski}, {Bolton},
  {Bovy}, {Shelden Bradley}, {Brandt}, {Brauer}, {Brinkmann}, {Brown},
  {Brownstein}, {Burden}, {Burtin}, {Busca}, {Cai}, {Capozzi}, {Carnero
  Rosell}, {Carr}, {Carrera}, {Chambers}, {Chaplin}, {Chen}, {Chiappini},
  {Chojnowski}, {Chuang}, {Clerc}, {Comparat}, {Covey}, {Croft}, {Cuesta},
  {Cunha}, {da Costa}, {Da Rio}, {Davenport}, {Dawson}, {De Lee}, {Delubac},
  {Deshpande}, {Dhital}, {Dutra-Ferreira}, {Dwelly}, {Ealet}, {Ebelke},
  {Edmondson}, {Eisenstein}, {Ellsworth}, {Elsworth}, {Epstein}, {Eracleous},
  {Escoffier}, {Esposito}, {Evans}, {Fan}, {Fern{\'a}ndez-Alvar}, {Feuillet},
  {Filiz Ak}, {Finley}, {Finoguenov}, {Flaherty}, {Fleming}, {Font-Ribera},
  {Foster}, {Frinchaboy}, {Galbraith-Frew}, {Garc{\'\i}a},
  {Garc{\'\i}a-Hern{\'a}ndez}, {Garc{\'\i}a P{\'e}rez}, {Gaulme}, {Ge},
  {G{\'e}nova-Santos}, {Georgakakis}, {Ghezzi}, {Gillespie}, {Girardi},
  {Goddard}, {Gontcho}, {Gonz{\'a}lez Hern{\'a}ndez}, {Grebel}, {Green},
  {Grieb}, {Grieves}, {Gunn}, {Guo}, {Harding}, {Hasselquist}, {Hawley},
  {Hayden}, {Hearty}, {Hekker}, {Ho}, {Hogg}, {Holley-Bockelmann}, {Holtzman},
  {Honscheid}, {Huber}, {Huehnerhoff}, {Ivans}, {Jiang}, {Johnson},
  {Kinemuchi}, {Kirkby}, {Kitaura}, {Klaene}, {Knapp}, {Kneib}, {Koenig},
  {Lam}, {Lan}, {Lang}, {Laurent}, {Le Goff}, {Leauthaud}, {Lee}, {Lee},
  {Licquia}, {Liu}, {Long}, {L{\'o}pez-Corredoira}, {Lorenzo-Oliveira},
  {Lucatello}, {Lundgren}, {Lupton}, {Mack}, {Mahadevan}, {Maia}, {Majewski},
  {Malanushenko}, {Malanushenko}, {Manchado}, {Manera}, {Mao}, {Maraston},
  {Marchwinski}, {Margala}, {Martell}, {Martig}, {Masters}, {Mathur},
  {McBride}, {McGehee}, {McGreer}, {McMahon}, {M{\'e}nard}, {Menzel},
  {Merloni}, {M{\'e}sz{\'a}ros}, {Miller}, {Miralda-Escud{\'e}}, {Miyatake},
  {Montero-Dorta}, {More}, {Morganson}, {Morice-Atkinson}, {Morrison},
  {Mosser}, {Muna}, {Myers}, {Nand ra}, {Newman}, {Neyrinck}, {Nguyen},
  {Nichol}, {Nidever}, {Noterdaeme}, {Nuza}, {O'Connell}, {O'Connell},
  {O'Connell}, {Ogando}, {Olmstead}, {Oravetz}, {Oravetz}, {Osumi}, {Owen},
  {Padgett}, {Padmanabhan}, {Paegert}, {Palanque-Delabrouille}, {Pan},
  {Parejko}, {P{\^a}ris}, {Park}, {Pattarakijwanich}, {Pellejero-Ibanez},
  {Pepper}, {Percival}, {P{\'e}rez-Fournon}, {Ṕrez-Ra`fols}, {Petitjean},
  {Pieri}, {Pinsonneault}, {Porto de Mello}, {Prada}, {Prakash},
  {Price-Whelan}, {Protopapas}, {Raddick}, {Rahman}, {Reid}, {Rich}, {Rix},
  {Robin}, {Rockosi}, {Rodrigues}, {Rodr{\'\i}guez-Torres}, {Roe}, {Ross},
  {Ross}, {Rossi}, {Ruan}, {Rubi{\~n}o-Mart{\'\i}n}, {Rykoff},
  {Salazar-Albornoz}, {Salvato}, {Samushia}, {S{\'a}nchez}, {Santiago},
  {Sayres}, {Schiavon}, {Schlegel}, {Schmidt}, {Schneider}, {Schultheis},
  {Schwope}, {Sc{\'o}ccola}, {Scott}, {Sellgren}, {Seo}, {Serenelli}, {Shane},
  {Shen}, {Shetrone}, {Shu}, {Silva Aguirre}, {Sivarani}, {Skrutskie},
  {Slosar}, {Smith}, {Sobreira}, {Souto}, {Stassun}, {Steinmetz}, {Stello},
  {Strauss}, {Streblyanska}, {Suzuki}, {Swanson}, {Tan}, {Tayar}, {Terrien},
  {Thakar}, {Thomas}, {Thomas}, {Thompson}, {Tinker}, {Tojeiro}, {Troup},
  {Vargas-Maga{\~n}a}, {Vazquez}, {Verde}, {Viel}, {Vogt}, {Wake}, {Wang},
  {Weaver}, {Weinberg}, {Weiner}, {White}, {Wilson}, {Wisniewski},
  {Wood-Vasey}, {Ye`che}, {York}, {Zakamska}, {Zamora}, {Zasowski}, {Zehavi},
  {Zhao}, {Zheng}, {Zhou}, {Zhou}, {Zou}, \& {Zhu}}]{2015ApJS..219...12A}
{Alam}, S., {Albareti}, F.~D., {Allende Prieto}, C., {et~al.} 2015, \apjs, 219,
  12

\bibitem[{{Aldering} {et~al.}(2002){Aldering}, {Adam}, {Antilogus}, {Astier},
  {Bacon}, {Bongard}, {Bonnaud}, {Copin}, {Hardin}, {Henault}, {Howell},
  {Lemonnier}, {Levy}, {Loken}, {Nugent}, {Pain}, {Pecontal}, {Pecontal},
  {Perlmutter}, {Quimby}, {Schahmaneche}, {Smadja}, \&
  {Wood-Vasey}}]{2002SPIE.4836...61A}
{Aldering}, G., {Adam}, G., {Antilogus}, P., {et~al.} 2002, Society of
  Photo-Optical Instrumentation Engineers (SPIE) Conference Series, Vol. 4836,
  {Overview of the Nearby Supernova Factory}, ed. J.~A. {Tyson} \& S.~{Wolff},
  61--72

\bibitem[{{Chambers} {et~al.}(2016){Chambers}, {Magnier}, {Metcalfe},
  {Flewelling}, {Huber}, {Waters}, {Denneau}, {Draper}, {Farrow}, {Finkbeiner},
  {Holmberg}, {Koppenhoefer}, {Price}, {Rest}, {Saglia}, {Schlafly}, {Smartt},
  {Sweeney}, {Wainscoat}, {Burgett}, {Chastel}, {Grav}, {Heasley}, {Hodapp},
  {Jedicke}, {Kaiser}, {Kudritzki}, {Luppino}, {Lupton}, {Monet}, {Morgan},
  {Onaka}, {Shiao}, {Stubbs}, {Tonry}, {White}, {Ba{\~n}ados}, {Bell},
  {Bender}, {Bernard}, {Boegner}, {Boffi}, {Botticella}, {Calamida},
  {Casertano}, {Chen}, {Chen}, {Cole}, {Deacon}, {Frenk}, {Fitzsimmons},
  {Gezari}, {Gibbs}, {Goessl}, {Goggia}, {Gourgue}, {Goldman}, {Grant},
  {Grebel}, {Hambly}, {Hasinger}, {Heavens}, {Heckman}, {Henderson}, {Henning},
  {Holman}, {Hopp}, {Ip}, {Isani}, {Jackson}, {Keyes}, {Koekemoer}, {Kotak},
  {Le}, {Liska}, {Long}, {Lucey}, {Liu}, {Martin}, {Masci}, {McLean}, {Mindel},
  {Misra}, {Morganson}, {Murphy}, {Obaika}, {Narayan}, {Nieto-Santisteban},
  {Norberg}, {Peacock}, {Pier}, {Postman}, {Primak}, {Rae}, {Rai}, {Riess},
  {Riffeser}, {Rix}, {R{\"o}ser}, {Russel}, {Rutz}, {Schilbach}, {Schultz},
  {Scolnic}, {Strolger}, {Szalay}, {Seitz}, {Small}, {Smith}, {Soderblom},
  {Taylor}, {Thomson}, {Taylor}, {Thakar}, {Thiel}, {Thilker}, {Unger},
  {Urata}, {Valenti}, {Wagner}, {Walder}, {Walter}, {Watters}, {Werner},
  {Wood-Vasey}, \& {Wyse}}]{2016arXiv161205560C}
{Chambers}, K.~C., {Magnier}, E.~A., {Metcalfe}, N., {et~al.} 2016, arXiv
  e-prints, arXiv:1612.05560

\bibitem[{{Courtois} \& {Tully}(2015)}]{2015MNRAS.447.1531C}
{Courtois}, H.~M. \& {Tully}, R.~B. 2015, \mnras, 447, 1531

\bibitem[{{Courtois} {et~al.}(2009){Courtois}, {Tully}, {Fisher}, {Bonhomme},
  {Zavodny}, \& {Barnes}}]{2009AJ....138.1938C}
{Courtois}, H.~M., {Tully}, R.~B., {Fisher}, J.~R., {et~al.} 2009, \aj, 138,
  1938

\bibitem[{{Courtois} {et~al.}(2017){Courtois}, {Tully}, {Hoffman},
  {Pomar{\`e}de}, {Graziani}, \& {Dupuy}}]{2017ApJ...847L...6C}
{Courtois}, H.~M., {Tully}, R.~B., {Hoffman}, Y., {et~al.} 2017, \apjl, 847, L6

\bibitem[{{Courtois} {et~al.}(2011){Courtois}, {Tully}, {Makarov}, {Mitronova},
  {Koribalski}, {Karachentsev}, \& {Fisher}}]{Courtois:2011ab}
{Courtois}, H.~M., {Tully}, R.~B., {Makarov}, D.~I., {et~al.} 2011, \mnras,
  414, 2005

\bibitem[{{Courtois} {et~al.}(2015){Courtois}, {Zaritsky}, {Sorce}, \&
  {Pomar{\`e}de}}]{Courtois:2015aa}
{Courtois}, H.~M., {Zaritsky}, D., {Sorce}, J.~G., \& {Pomar{\`e}de}, D. 2015,
  \mnras, 448, 1767

\bibitem[{{Giovanelli} {et~al.}(2007){Giovanelli}, {Haynes}, {Kent},
  {Saintonge}, {Stierwalt}, {Altaf}, {Balonek}, {Brosch}, {Brown}, {Catinella},
  {Furniss}, {Goldstein}, {Hoffman}, {Koopmann}, {Kornreich}, {Mahmood},
  {Martin}, {Masters}, {Mitschang}, {Momjian}, {Nair}, {Rosenberg}, \&
  {Walsh}}]{2007AJ....133.2569G}
{Giovanelli}, R., {Haynes}, M.~P., {Kent}, B.~R., {et~al.} 2007, \aj, 133, 2569

\bibitem[{{Haynes} {et~al.}(2018){Haynes}, {Giovanelli}, {Kent}, {Adams},
  {Balonek}, {Craig}, {Fertig}, {Finn}, {Giovanardi}, {Hallenbeck}, {Hess},
  {Hoffman}, {Huang}, {Jones}, {Koopmann}, {Kornreich}, {Leisman}, {Miller},
  {Moorman}, {O'Connor}, {O'Donoghue}, {Papastergis}, {Troischt}, {Stark}, \&
  {Xiao}}]{2018ApJ...861...49H}
{Haynes}, M.~P., {Giovanelli}, R., {Kent}, B.~R., {et~al.} 2018, \apj, 861, 49

\bibitem[{{Haynes} {et~al.}(2011){Haynes}, {Giovanelli}, {Martin}, {Hess},
  {Saintonge}, {Adams}, {Hallenbeck}, {Hoffman}, {Huang}, {Kent}, {Koopmann},
  {Papastergis}, {Stierwalt}, {Balonek}, {Craig}, {Higdon}, {Kornreich},
  {Miller}, {O'Donoghue}, {Olowin}, {Rosenberg}, {Spekkens}, {Troischt}, \&
  {Wilcots}}]{2011AJ....142..170H}
{Haynes}, M.~P., {Giovanelli}, R., {Martin}, A.~M., {et~al.} 2011, \aj, 142,
  170

\bibitem[{{Hoffman} {et~al.}(2017){Hoffman}, {Pomar{\`e}de}, {Tully}, \&
  {Courtois}}]{Hoffman:2017aa}
{Hoffman}, Y., {Pomar{\`e}de}, D., {Tully}, R.~B., \& {Courtois}, H.~M. 2017,
  Nature Astronomy, 1, 0036

\bibitem[{{Hong} {et~al.}(2013){Hong}, {Staveley-Smith}, {Masters}, {Springob},
  {Macri}, {Koribalski}, {Jones}, {Jarrett}, \& {Crook}}]{2013MNRAS.432.1178H}
{Hong}, T., {Staveley-Smith}, L., {Masters}, K.~L., {et~al.} 2013, \mnras, 432,
  1178

\bibitem[{{Huchtmeier} {et~al.}(2005){Huchtmeier}, {Karachentsev},
  {Karachentseva}, {Kudrya}, \& {Mitronova}}]{2005A&A...435..459H}
{Huchtmeier}, W.~K., {Karachentsev}, I.~D., {Karachentseva}, V.~E., {Kudrya},
  Y.~N., \& {Mitronova}, S.~N. 2005, \aap, 435, 459

\bibitem[{{Hunt} {et~al.}(2016){Hunt}, {Pisano}, \&
  {Edel}}]{2016AJ....152...30H}
{Hunt}, L.~R., {Pisano}, D.~J., \& {Edel}, S. 2016, \aj, 152, 30

\bibitem[{{Karachentsev} {et~al.}(1999){Karachentsev}, {Karachentseva},
  {Kudrya}, {Sharina}, \& {Parnovskij}}]{Karachentsev:1999aa}
{Karachentsev}, I.~D., {Karachentseva}, V.~E., {Kudrya}, Y.~N., {Sharina},
  M.~E., \& {Parnovskij}, S.~L. 1999, Bulletin of the Special Astrophysics
  Observatory, 47 [\eprint{astro-ph/0305566}]

\bibitem[{{Kent} {et~al.}(2008){Kent}, {Giovanelli}, {Haynes}, {Martin},
  {Saintonge}, {Stierwalt}, {Balonek}, {Brosch}, \&
  {Koopmann}}]{2008AJ....136..713K}
{Kent}, B.~R., {Giovanelli}, R., {Haynes}, M.~P., {et~al.} 2008, \aj, 136, 713

\bibitem[{{Koribalski} {et~al.}(2004){Koribalski}, {Staveley-Smith}, {Kilborn},
  {Ryder}, {Kraan-Korteweg}, {Ryan-Weber}, {Ekers}, {Jerjen}, {Henning},
  {Putman}, {Zwaan}, {de Blok}, {Calabretta}, {Disney}, {Minchin}, {Bhathal},
  {Boyce}, {Drinkwater}, {Freeman}, {Gibson}, {Green}, {Haynes}, {Juraszek},
  {Kesteven}, {Knezek}, {Mader}, {Marquarding}, {Meyer}, {Mould}, {Oosterloo},
  {O'Brien}, {Price}, {Sadler}, {Schr{\"o}der}, {Stewart}, {Stootman}, {Waugh},
  {Warren}, {Webster}, \& {Wright}}]{2004AJ....128...16K}
{Koribalski}, B.~S., {Staveley-Smith}, L., {Kilborn}, V.~A., {et~al.} 2004,
  \aj, 128, 16

\bibitem[{{Kourkchi} {et~al.}(2020{\natexlab{a}}){Kourkchi}, {Courtois},
  {Graziani}, {Hoffman}, {Pomar{\`e}de}, {Shaya}, \&
  {Tully}}]{2020AJ....159...67K}
{Kourkchi}, E., {Courtois}, H.~M., {Graziani}, R., {et~al.} 2020{\natexlab{a}},
  \aj, 159, 67

\bibitem[{{Kourkchi} {et~al.}(2020{\natexlab{b}}){Kourkchi}, {Tully},
  {Eftekharzadeh}, {Llop}, {Courtois}, {Guinet}, {Dupuy}, {Neill}, {Seibert},
  {Andrews}, {Chuang}, {Danesh}, {Gonzalez}, {Holthaus}, {Mokelke}, {Schoen},
  \& {Urasaki}}]{2020ApJ...902..145K}
{Kourkchi}, E., {Tully}, R.~B., {Eftekharzadeh}, S., {et~al.}
  2020{\natexlab{b}}, \apj, 902, 145

\bibitem[{{Kourkchi} {et~al.}(2019){Kourkchi}, {Tully}, {Neill}, {Seibert},
  {Courtois}, \& {Dupuy}}]{2019ApJ...884...82K}
{Kourkchi}, E., {Tully}, R.~B., {Neill}, J.~D., {et~al.} 2019, \apj, 884, 82

\bibitem[{Luo {et~al.}(2015)Luo, Zhao, Zhao, Deng, Liu, Jing, Wang, Zhang, Shi,
  Cui, Chu, Li, Bai, Wu, Cai, Cao, Cao, Carlin, Chen, Chen, Chen, Chen, Chen,
  Chen, Chen, Christlieb, Chu, Cui, Dong, Du, Fan, Feng, Fu, Gao, Gong, Gu,
  Guo, Han, He, Hou, Hou, Hou, Hu, Hu, Hu, Huo, Jia, Jiang, Jiang, Jiang, Jin,
  Kong, Kong, Lei, Li, Li, Li, Li, Li, Li, Li, Li, Li, Li, Li, Li, Liang, Lin,
  Liu, Liu, Liu, Liu, Lu, Luo, Mao, Newberg, Ni, Qi, Qi, Shen, Shi, Song, Song,
  Su, Su, Tang, Tao, Tian, Wang, Wang, Wang, Wang, Wang, Wang, Wang, Wang,
  Wang, Wang, Wang, Wang, Wang, Wang, Wang, Wang, Wang, Wang, Wang, Wang, Wei,
  Wei, Wu, Wu, Wu, Wu, Xing, Xu, Xu, Xu, Yan, Yang, Yang, Yang, Yang, Yao, Yu,
  Yuan, Yuan, Yuan, Yuan, Zhai, Zhang, Zhang, Zhang, Zhang, Zhang, Zhang,
  Zhang, Zhang, Zhao, Zhou, Zhou, Zhu, Zhu, Zou, \& Zuo}]{Luo_2015}
Luo, A.-L., Zhao, Y.-H., Zhao, G., {et~al.} 2015, Research in Astronomy and
  Astrophysics, 15, 1095

\bibitem[{{Makarov} {et~al.}(2014){Makarov}, {Prugniel}, {Terekhova},
  {Courtois}, \& {Vauglin}}]{2014A&A...570A..13M}
{Makarov}, D., {Prugniel}, P., {Terekhova}, N., {Courtois}, H., \& {Vauglin},
  I. 2014, \aap, 570, A13

\bibitem[{{Rigault} {et~al.}(2020){Rigault}, {Brinnel}, {Aldering},
  {Antilogus}, {Aragon}, {Bailey}, {Baltay}, {Barbary}, {Bongard}, {Boone},
  {Buton}, {Childress}, {Chotard}, {Copin}, {Dixon}, {Fagrelius}, {Feindt},
  {Fouchez}, {Gangler}, {Hayden}, {Hillebrandt}, {Howell}, {Kim}, {Kowalski},
  {Kuesters}, {Leget}, {Lombardo}, {Lin}, {Nordin}, {Pain}, {Pecontal},
  {Pereira}, {Perlmutter}, {Rabinowitz}, {Runge}, {Rubin}, {Saunders},
  {Smadja}, {Sofiatti}, {Suzuki}, {Taubenberger}, {Tao}, \&
  {Thomas}}]{2018arXiv180603849R}
{Rigault}, M., {Brinnel}, V., {Aldering}, G., {et~al.} 2020, \aap, 644, A176

\bibitem[{{Saintonge} {et~al.}(2008){Saintonge}, {Giovanelli}, {Haynes},
  {Hoffman}, {Kent}, {Martin}, {Stierwalt}, \& {Brosch}}]{2008AJ....135..588S}
{Saintonge}, A., {Giovanelli}, R., {Haynes}, M.~P., {et~al.} 2008, \aj, 135,
  588

\bibitem[{{Springob} {et~al.}(2005){Springob}, {Haynes}, {Giovanelli}, \&
  {Kent}}]{2005ApJS..160..149S}
{Springob}, C.~M., {Haynes}, M.~P., {Giovanelli}, R., \& {Kent}, B.~R. 2005,
  \apjs, 160, 149

\bibitem[{{Theureau} {et~al.}(1998){Theureau}, {Bottinelli}, {Coudreau-Durand},
  {Gouguenheim}, {Hallet}, {Loulergue}, {Paturel}, \&
  {Teerikorpi}}]{Theureau:1998a}
{Theureau}, G., {Bottinelli}, L., {Coudreau-Durand}, N., {et~al.} 1998, \aaps,
  130, 333

\bibitem[{{Theureau} {et~al.}(2005){Theureau}, {Coudreau}, {Hallet}, {Hanski},
  {Alsac}, {Bottinelli}, {Gouguenheim}, {Martin}, \&
  {Paturel}}]{Theureau:2005aa}
{Theureau}, G., {Coudreau}, N., {Hallet}, N., {et~al.} 2005, \aap, 430, 373

\bibitem[{{Theureau} {et~al.}(2017){Theureau}, {Coudreau}, {Hallet}, {Hanski},
  \& {Poulain}}]{Theureau:2017aa}
{Theureau}, G., {Coudreau}, N., {Hallet}, N., {Hanski}, M.~O., \& {Poulain}, M.
  2017, \aap, 599, A104

\bibitem[{{Theureau} {et~al.}(2007){Theureau}, {Hanski}, {Coudreau}, {Hallet},
  \& {Martin}}]{2007A&A...465...71T}
{Theureau}, G., {Hanski}, M.~O., {Coudreau}, N., {Hallet}, N., \& {Martin},
  J.~M. 2007, \aap, 465, 71

\bibitem[{{Theureau} {et~al.}(2006){Theureau}, {Martin}, {Cognard}, \&
  {Borsenberger}}]{2006ASPC..351..429T}
{Theureau}, G., {Martin}, J.~M., {Cognard}, I., \& {Borsenberger}, J. 2006, in
  Astronomical Society of the Pacific Conference Series, Vol. 351, Astronomical
  Data Analysis Software and Systems XV, ed. C.~{Gabriel}, C.~{Arviset},
  D.~{Ponz}, \& S.~{Enrique}, 429

\bibitem[{{Tully} {et~al.}(2014){Tully}, {Courtois}, {Hoffman}, \&
  {Pomar{\`e}de}}]{2014Natur.513...71T}
{Tully}, R.~B., {Courtois}, H., {Hoffman}, Y., \& {Pomar{\`e}de}, D. 2014,
  \nat, 513, 71

\bibitem[{{Tully} \& {Fouque}(1985)}]{1985ApJS...58...67T}
{Tully}, R.~B. \& {Fouque}, P. 1985, \apjs, 58, 67

\bibitem[{{Tully} {et~al.}(2019){Tully}, {Pomar{\`e}de}, {Graziani},
  {Courtois}, {Hoffman}, \& {Shaya}}]{2019ApJ...880...24T}
{Tully}, R.~B., {Pomar{\`e}de}, D., {Graziani}, R., {et~al.} 2019, \apj, 880,
  24

\bibitem[{{van Driel} {et~al.}(2016){van Driel}, {Butcher}, {Schneider},
  {Lehnert}, {Minchin}, {Blyth}, {Chemin}, {Hallet}, {Joseph}, {Kotze},
  {Kraan-Korteweg}, {Olofsson}, \& {Ramatsoku}}]{2016A&A...595A.118V}
{van Driel}, W., {Butcher}, Z., {Schneider}, S., {et~al.} 2016, \aap, 595, A118

\bibitem[{{van Driel} {et~al.}(2008){van Driel}, {Schneider}, {Lehnert}, \&
  {Minchin}}]{van-Driel:2008aa}
{van Driel}, W., {Schneider}, S.~E., {Lehnert}, M., \& {Minchin}, R. 2008, in
  American Institute of Physics Conference Series, Vol. 1035, The Evolution of
  Galaxies Through the Neutral Hydrogen Window, ed. R.~{Minchin} \&
  E.~{Momjian}, 256--258

\bibitem[{{Zwaan} {et~al.}(2001){Zwaan}, {Briggs}, \&
  {Sprayberry}}]{2001MNRAS.327.1249Z}
{Zwaan}, M.~A., {Briggs}, F.~H., \& {Sprayberry}, D. 2001, \mnras, 327, 1249

\end{thebibliography}

\end{document}